\documentclass[11pt]{article}
\pdfoutput=1  
\usepackage{jcappub}
\usepackage{graphicx}
\usepackage{amsmath,amssymb}
\graphicspath{{./figures/}}
\usepackage{appendix}
\usepackage{placeins}
\usepackage[normalem]{ulem}
\usepackage[utf8]{inputenc}

\input epsf
\usepackage{mathtools}
\usepackage{amsfonts}
\usepackage{upgreek}
\usepackage{latexsym}
\usepackage{stfloats}
\usepackage{afterpage}

\newcommand{\be}{\begin{equation}}
\newcommand{\ee}{\end{equation}}
\newcommand{\beq}{\begin{equation}}
\newcommand{\eeq}{\end{equation}}
\newcommand{\bea}{\begin{eqnarray}}
\newcommand{\eea}{\end{eqnarray}}


\newcommand{\dd}{\text{d}}
\newcommand{\bx}{{\bf{x}}}
\newcommand{\by}{{\bf{y}}}

\newcommand{\De}{\Delta}

\newcommand{\nn}{{\nonumber}}

\newcommand{\NN}{{\cal N}}

\usepackage[stable]{footmisc}

\newcommand{\al}{\alpha}
\newcommand{\bal}{{\boldsymbol{\alpha}}}

\newcommand{\bbt}{{\boldsymbol{\beta}}}
\newcommand{\bth}{{\boldsymbol{\theta}}}
\newcommand{\bxi}{{\boldsymbol{\xi}}}

\newcommand{\ep}{\epsilon}
\newcommand{\ka}{\kappa}

\renewcommand{\th}{\theta}

\newcommand{\si}{\sigma}
\newcommand{\Si}{\Sigma}

\newcommand{\ra}{\rightarrow}
\newcommand{\bnabla}{{\boldsymbol{\nabla}}}
\newcommand\ees{\end{eqnarray}}
\newcommand\bees{\begin{eqnarray}}

 \definecolor{magenta}{rgb}{0.1,0.98,0.6}

\definecolor{dgreen}{rgb}{0,0.7,0.0}

\title{Strong and weak lensing of Gravitational Waves: a semi-analytical approach}
\author{Giulia Cusin$^a$}
\affiliation{$^a$Astrophysics, University of Oxford, DWB, Keble Road, Oxford, OX1 3RH, UK} 
\author{Ruth Durrer$^b$}
\affiliation{$^b$Universit\'e de Gen\`eve, D\'epartement de Physique Th\'eorique and Centre for Astroparticle Physics,
24 quai Ernest-Ansermet, CH-1211 Gen\`eve 4, Switzerland}
\author{Irina Dvorkin$^c$}
\affiliation{$^c$Sorbonne Universit\'{e}, CNRS, UMR 7095, Institut d’Astrophysique de Paris, 98 bis boulevard Arago, 75014 Paris,France}

\emailAdd{giulia.cusin@physics.ox.ac.uk}
\emailAdd{ruth.durrer@unige.ch}
\emailAdd{dvorkin@iap.fr}

\abstract{
In this paper we study gravitational lensing of gravitational wave events. The main point of the present work is to introduce a semi-analytic approach so that each ingredient can be varied and tested individually. Our analytic model for the source population is motivated by a numerical study and we compare semi-analytical and numerical results. We determine the expected magnification for events seen at a given observed luminosity distance. We find that while the probability of significant magnification of the observed LIGO-Virgo events is small in our model, the probability distribution of the magnification has a  significant tail to high $\mu$ such that e.g. the variance of the magnification is very large and even diverges in the geometric optics approximation. For the 10 binary black hole mergers observed by LIGO-Virgo in the O1+O2 observation campaigns, with our modelling we find that the probability that one of them has been magnified with magnification of 5 or bigger is $\mathcal{P}_{\text{obs}}(>5)\sim 0.01$ while the probability  of magnification by 50 or bigger is $\mathcal{P}_{\text{obs}}(>50)\sim 0.005$.%
}

\begin{document}

\maketitle

\section{Introduction}
The direct detection of gravitational waves from inspiraling binary black holes and neutron stars by the LIGO and Virgo experiments~\cite{Abbott:2016blz,Abbott:2016nmj,TheLIGOScientific:2016pea,Abbott:2017gyy,Abbott:2017oio,Abbott:2017vtc,TheLIGOScientific:2017qsa,LIGOScientific:2018mvr} has been one of the greatest successes in experimental physics and observational astronomy in recent years.  Not only does it lead to unprecedented insights on black holes and especially neutron stars, but it also allows us to test general relativity (GR) and cosmology.

An interesting aspect is the fact that the clustered matter between the gravitational wave source and the observer can  amplify the observed signal. This can lead to an underestimation of the distance to the source and therefore  of its redshift which in turn leads  to an overestimation of the chirp mass~\cite{Oguri:2018muv}. The effects of gravitational lensing on the apparent properties of detected sources, and possible identification of unique lensing features were extensively studied~\cite{Wang_1996, Sereno_2010,Ng:2017yiu,Broadhurst:2018saj,Broadhurst:2019ijv,Haris:2018vmn,Cao:2014oaa,Dai:2017huk,Dai:2018enj,Dai:2016igl,Jung:2017flg,Lai:2018rto,Smith:2017mqu} with the general conclusion that such effects can become important when current detectors (LIGO/Virgo) achieve their design sensitivities, and especially in the era of third generation detectors (Einstein Telescope, Cosmic Explorer).

In this paper, we want to determine the probability that an event of a given strain amplitude, or rather of a given signal to noise ratio (SNR) for fixed experiment characteristics, is being significantly magnified by lensing. We want to study the dependence of the strong lensing probability and of the mean magnification on redshift and on the strain amplitude (or SNR). 
Similar studies are already present in the literature~\cite{Smith:2017mqu,Ng:2017yiu,Oguri:2018muv,Hannuksela:2019kle,Oguri:2019fix,Diego:2019rzc,Haris:2018vmn}. The main difference to the present work is that instead of full numerical simulations, we present semi-analytic formulae which can be applied to an arbitrary lens and source distribution keeping full control of modelling and transparency of all physical effects. This allows us to study the dependence of the result on the lens and/or source distribution which otherwise is quite obscure.
We model the lenses as singular isothermal spheres (SIS) which is reasonable for galaxies. While SIS are not  realistic when considering lensing by individual clusters, they are roughly sufficient for statistical purposes, see ~\cite{Gavazzi:2007vw,Robertson:2020mfh} for a detailed study. The main advantage of SIS is of course that they allow for analytic results.

We concentrate on short-lived sources, such as compact binary coalescences which can be observed with LIGO-Virgo \cite{Abbott:2016blz,Abbott:2016nmj,Abbott:2017vtc,TheLIGOScientific:2016pea,Abbott:2017oio,Abbott:2017gyy,LIGOScientific:2018mvr}. 
Interestingly, the first detections of binary black holes (BBHs) revealed a population of sources more massive than those known from X-ray binaries, with masses of up to $50M_{\odot}$ discovered during the second LIGO observation run \cite{Abbott:2017vtc,LIGOScientific:2018mvr}. Such high masses are in fact predicted by various stellar evolution models \cite[e.g.]{Belczynski:2016obo,Eldridge:2016ymr,Wysocki:2017isg,Dvorkin:2017kfg,Kruckow:2018slo,Spera:2018wnw,Mapelli:2019bnp,Rodriguez:2019huv,Stevenson:2019rcw,Fragione:2019vgr}. Nevertheless, these findings inspired the authors of Ref. \cite{Broadhurst:2018saj} to suggest that the intrinsic masses of the observed BBHs were much lower, closer to the ones found in X-ray binaries, and the apparent high masses are the result of redshifting, implying that these are high-redshift sources that underwent gravitational lensing. For this, however, the authors have to assume a high lensing probability which, as we shall see in this paper, cannot be obtained within the standard cosmological model. The detailed analysis of the individual events in Ref. \cite{Hannuksela:2019kle} found that the probability of lensing for the sources observed by LIGO-Virgo so far is vanishingly small, with the consequence that BH population models can be reasonably informed by GW observations to date, as summarized in the first GW source catalogue \cite{LIGOScientific:2018mvr,LIGOScientific:2018jsj}.
In this work we shall confirm this finding with a statistical analysis. We present predictions for the probability that an event observed by LIGO-Virgo at a given distance has been magnified by a given amount. We find that while the probability is very small, it decays very slowly with increasing magnification, i.e. there is a non-negligible probability of having extremely highly magnified events. 

The remainder of this paper is structured as follows: in the next section we specify our notation and present a glossary with acronyms and definitions. We then present the lens model and we derive expressions for the cross section, optical depth and lensing probability. In section \ref{e:stat} we derive the expression for the probability that a source seen from a given \emph{observed} distance has been magnified by a given factor and we compute the mean magnification and its variance both as a function of the (unobservable)  cosmological redshift and the observed one. In section~\ref{s:ana} we discuss a simple analytical model which is useful to understand the physics and we present results for the mean magnification for different choices of the source distribution. In section~\ref{s:num} we introduce a numerical study for the population of lenses and sources and  we present numerical results for the magnification as a function of both cosmological and observed redshift, for the LIGO-Virgo detector network. We also compute the probability of magnification for a source with given observed properties. In section~\ref{s:con} we discuss our results and conclude. Some details are deferred to multiple appendices.

\section{Lens modeling}\label{s:theo}

In this section we present our theoretical model: we describe galaxies acting as lenses as singular isothermal spheres (SIS) and we derive expressions for the magnification and time delay between multiple images as a function of the geometrical quantities entering the picture.  SIS are not a very realistic lens model but since we are interested in statistical averages and not in the detailed modelling of a single lens, the fact that individual lenses will in general be triaxial is expected to average out. Notice that substructure may actually lead to additional magnification. In this sense our model is rather conservative. However, a recent numerical study has shown that the SIS model is surprisingly close to 
the observed lensing probability~\cite{Robertson:2020mfh}. Note also, that our results are only sensitive to the magnification of the most highly magnified image but not to the geometry or the number of images.

\subsection{Notation}
Before we start out theoretical description, we fix the notation used in this paper and present a glossary of the variables which will be used throughout.

\begin{figure}[ht!]
\begin{center}
\includegraphics[width=0.65\textwidth]{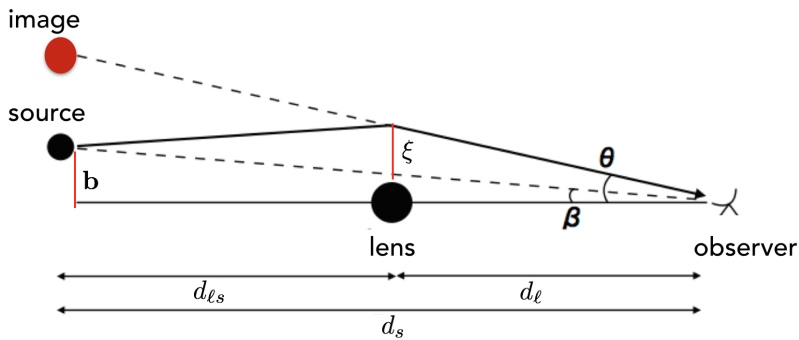}
\end{center}
\caption{\label{f:lens} A gravitational lens.} \end{figure}

In this paper we set $c=1$ but keep the gravitational constant $G$.  In Fig.\,\ref{f:lens} we schematically illustrate a lens. Standard textbook results (see e.g.~\cite{1992grle.book}) on strong lensing quantities which we use in this article are summarized in appendix \ref{AppA}. 
 We define  $\bth$ and $\bbt$ as the lensed and the unlensed angular position of the source. The optical axis is connecting the observer to the center of mass of the lens. We introduce  a vector in the source plane from the optical axis to the source,  ${\bf{b}}=d_s \bbt$ 
and a vector $\bxi$ in the lens plane defined as  $\bxi=\bth d_{\ell}$. This fixes the geometry. Here $d_{\ell}=d(z_{\ell})\,,~ d_{\ell s}=d(z_{\ell},z_s)$ and $d_s=d(z_s)$ are the angular diameter distances from the observer to the lens, from the lens to the source and from the observer to the source respectively.
The angular diameter distances in a spatially flat Friedman universe are defined as 
\bea\label{def}
d_\ell= d(z_{\ell}) &=& \frac{1}{1+z_{\ell}}\chi(0,z_{\ell}) \,,\\
d_{\ell s}=d(z_{\ell},z_s) &=&  \frac{1}{1+z_s}\chi(z_{\ell},z_s) \,,\\
d_s=d(z_s) &=&  \frac{1}{1+z_s}\chi(0,z_s)\,,\quad \mbox{ and}\\
\chi(z_1,z_2) &=& \int_{z_1}^{z_2}\frac{dz}{H(z)}=\chi(z_2)-\chi(z_1)\,, \quad \Big(~\chi(0,z)\equiv \chi(z)~\Big)
\eea
is the comoving distance from redshift $z_1$ to $z_2$. Here we assume vanishing spatial curvature and we set the present scale factor, $a_0=1$. 
In the standard $\Lambda$CDM cosmology the  angular diameter distance grows until about redshift $z=1.6$ after which it decays and tends to zero roughly like $1/(1+z)$.

A glossary with a list of the main symbols we will introduce in this article and their definition can be found in Table\,\ref{glossary}. 

\begin{table}[ht!]
\small
\begin{tabular}{c|c}
\hline
{\bf{symbol}}& {\bf{definition}}\\
 \hline\\
 $z_s$ & (cosmological) source  redshift \\
 $z_{\ell}$ &  (cosmological) lens redshift \\ 
 $D_s$  & luminosity distance to the source \\
 $D_{\ell}$ &luminosity distance to the lens   \\
$d_s$ &  angular diameter distance to the source\\
$d_{\ell}$ &  angular diameter distance to the lens\\
$d_{\ell s}$ &  angular diameter distance from the lens to the source\\
 $z_{\text{obs}}$ & observed source redshift (inferred assuming no magnification)\\
$\sigma_v$ &  galaxy velocity dispersion \\
$\mu$ & magnification \\
$\Delta t$ & time-delay between multiple images \\
$\sigma$ &  lensing cross-section\\
$\tau(\mu, z_s)$ & optical depth for magnification of at least $\mu$\\
$p(\mu, z_s)$ & probability density of magnification for source at $z_s$\\
$P(>\mu, z_s)$ & probability  of magnification of at least $\mu$ for source at $z_s$\\
 $F_{\text{lim}}$ & limiting flux of a given observatory\\
$d\mathcal{N}_{\text{obs}}(F_{\text{lim}}, z_s)$ & \# sources observed with liming flux $F_{\text{lim}}$ in a bin around $z_s$\\
$d\mathcal{N}_{\text{obs}}(F_{\text{lim}}, D_{\text{obs}})$ & \# sources observed with liming flux $F_{\text{lim}}$ in a bin around $D_{\text{obs}}$\\
$d\mathcal{N}(\mu, F_{\text{lim}}, z_s)$ & \# sources that we see with liming flux $F_{\text{lim}}$ in a bin around $z_s$, if the magnification is $\mu$ \\
$d\mathcal{N}(\mu, F_{\text{lim}}, D_{\text{obs}})$ & \# sources that we see with liming flux $F_{\text{lim}}$ in a bin around $D_{\text{obs}}$, if the magnification is $\mu$ \\
$\mathcal{P}_{z}(\mu, F_{\text{lim}})$ & probability density that an \emph{observed} source at (cosmological) redshift $z$ is magnified by $\mu$ \\
$\mathcal{P}_{\text{obs}}(\mu, F_{\text{lim}})$ & probability density that an \emph{observed} source at observed redshift $z_{\text{obs}}$ is magnified by $\mu$ \\
&\\
\hline
\end{tabular}
\caption{\label{glossary} \normalsize Glossary of the main quantities which we will introduce and  use in this paper.}
\end{table}

\subsection{Singular isothermal sphere: amplification and time delay}

 The matter density of a \emph{singular isothermal sphere} is given by
\be
\rho(r) = \frac{\si^2_v}{2\pi Gr^2} \,,
\ee
where $\si_v^2$ denotes the velocity dispersion of the lens.
Despite its singularity in the center  and infinite total mass, this can be  considered as a rather realistic mass distribution for lensing by a galaxy (see~\cite{1992grle.book,Oguri:2019fix});  $\sigma_v$ is the velocity dispersion inside the galaxy. 
Integrating along the line of sight we obtain
\be
\Si(\th) = \frac{\si_v^2}{2G}\frac{1}{d_{\ell}\th} \,.
\ee
We introduce $\Si_c=(4\pi G)^{-1} d_s/(d_{\ell s}d_{\ell})$ so that
\be
\ka(\th) = \frac{\Si(\th) }{\Si_c} = 2\pi\si_v^2\frac{d_{\ell s}}{d_s}\frac{1}{\th} \,.
\ee
A SIS has a constant  deflection angle given by 
\be
\al(\th) =4\pi\si_v^2\frac{d_{\ell s}}{d_s} \equiv \al_0\,.
\ee
This can be verified by inserting $\ka$ in the general expression for the 2-dimensional deflection angle \eqref{e:al1}. 
We now rescale our variables by $\al_0$, defining the rescaled image and source positions as $\bx=\bth/\al_0$ and $\by={\bbt}/\al_0$, respectively. 
In terms of these rescaled variables, the lens equation for the SIS simply becomes
\be
\by =\bx -\frac{\bx}{|\bx|} \,.
\ee
For $\by=0$ the solution is the Einstein ring $|\bx|=1$, for $y=|\by|<1$ there are two solutions. One with $x_1=|\bx_1|=1+y$ on the same side of the line of sight as the source (positive parity)  and one with  $x_2=|\bx_2|=1-y$ on the opposite side (negative parity). For $y>1$ the second solution no longer exists.
The Jacobian of the lens map is
\be
A_{ij} = \delta_{ij}\left(1- \frac{1}{|\bx|}\right) +\frac{x_ix_j}{|\bx|^3}\,,\quad {\rm det}A= 1- \frac{1}{|\bx|}\,, \qquad \mu = \frac{1}{|{\rm det}A|} = \frac{|\bx|}{|1-|\bx||} \,,
\ee
where we have introduced the magnification  $\mu$.
Expressing the total magnification of a point source at position $y$ in terms of $y$ we find
\be\label{e:muy}
\mu(y) = \left\{\begin{array}{ll} \mu(\bx_1) + \mu(\bx_2) =\frac{y+1}{y} +\frac{1-y}{y} =\mu_1+\mu_2=\frac{2}{y} \,, & y\leq1\,,\\
 \frac{y+1}{y} =1+\frac{1}{y} \,, & y\geq 1\,. \end{array} \right.
\ee
The lensing potential for the SIS is simply $\psi(x)=|x|$ and the time delay between the two images is given by
\be\label{Deltat}
\Delta t=\left[4\pi \sigma_v^2\right]^2 \frac{d_{\ell} d_{\ell s}}{d_s}(1+z_{\ell})2 y \equiv\beta_t(\si_v^2,z_\ell,z_s) y\,, 
\ee
where $2 y=x_1-x_2$.

\subsection{Cross section}

\begin{figure}[htt!]
\begin{center}
\includegraphics[width=0.45\textwidth]{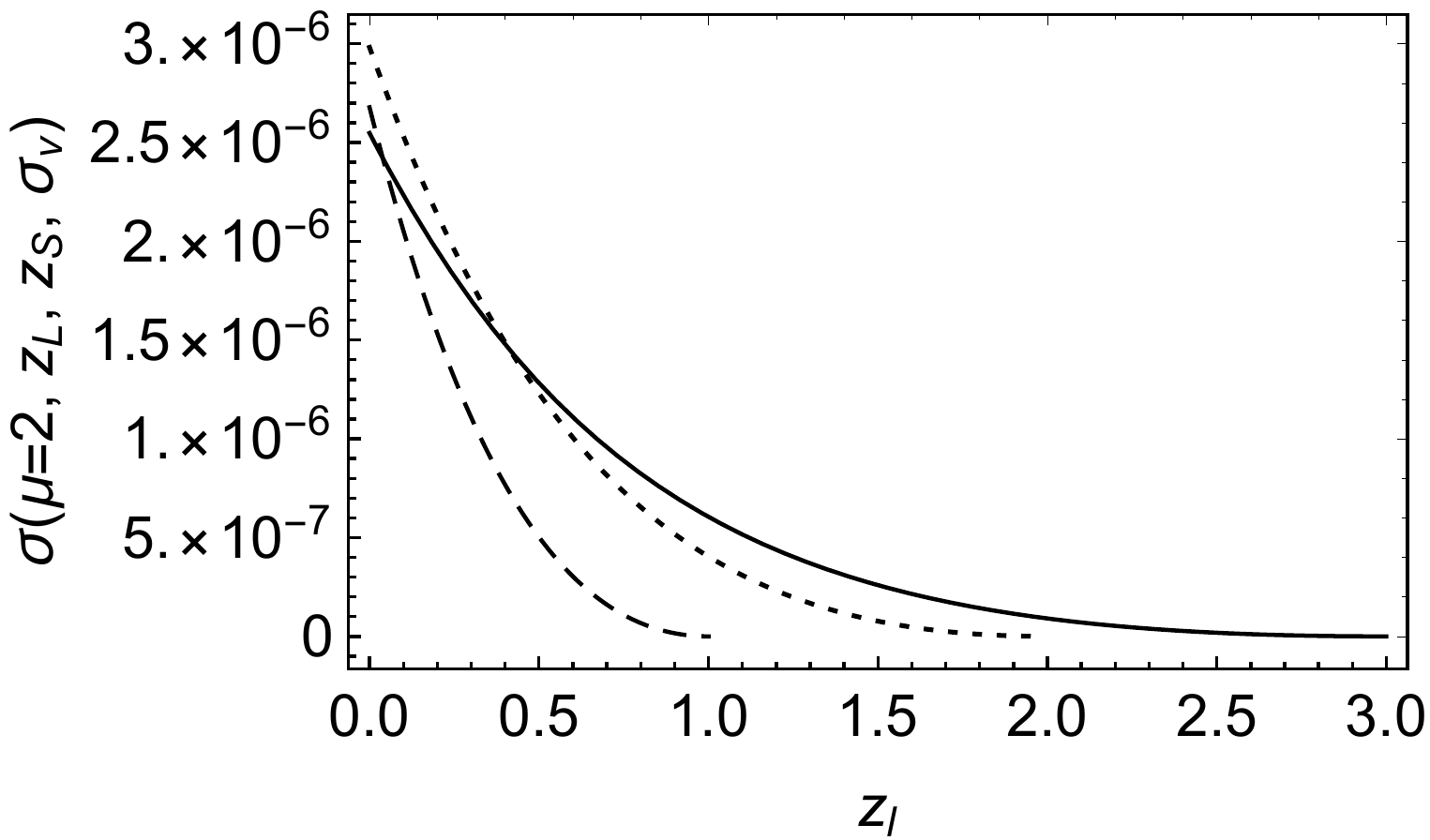} 
 \end{center}
\caption{\label{f:cross} Cross section (in units of Mpc$^2$) for $\mu_1=2$ and $\sigma_v=63$ km/s, in a $\Lambda$CDM universe. Different curves are for different source redshifts:  $z_s=1, 2, 3$ (dashed, dotted and continuous respectively). }
\end{figure}

The impact parameter of the source (in the source plane) is given by $|{\bf{b}}|=b=\beta d_s=y\al_0d_s$. A source with impact parameter smaller or equal to $b$ is amplified by at least a factor $\mu(y)$. Hence the cross section for amplification by  more than a factor $\mu_1$ of the stronger image from a SIS lens with velocity dispersion $\si_v$ is
\be\label{e:cross}
\sigma(\mu_1,z_{\ell},z_s,\si_v)= \pi b^2 =\pi(y\al_0d_s)^2 = \frac{\pi(4\pi)^2\si_v^4d_{\ell s}^2}{(\mu_1-1)^2}\,.
\ee
Note that this cross section gives the area, centered along the line of sight of the lens,  within which a source at $z_s$ must lie so that it is amplified by a factor $\mu$ or larger by the lens at $z_\ell$. 
The expression (\ref{e:cross}) remains valid also for $y\geq1$, where we have only one image with magnification $\mu_1$ which tends to $1$ when $y\ra\infty$.  In Fig.\,\ref{f:cross} we show the  cross-section given in eq.~\eqref{e:cross}, as a function of the lens redshift and for different redshifts of the GW source. Note also that the cross-section is by definition a positive definite quantity, however in a $\Lambda$CDM universe it is not monotonically increasing with $z_s$. 
In Fig.~\ref{f:cross} we see that for $z_\ell<0.4$,  the cross section $\sigma(\mu_1,z_{\ell},2,\si_v)>\sigma(\mu_1,z_{\ell},3,\si_v)$ and at very low redshift of the lens $z_{\ell}$ even $\sigma(\mu_1,z_{\ell},1,\si_v)>\sigma(\mu_1,z_{\ell},3,\si_v)$.  Actually, at fixed $z_\ell$, the area distance $d_{\ell s}$  and therefore also $\si(\mu,z_\ell,z_s)$  starts decreasing roughly when $z_s\gtrsim 2.2z_\ell+1.5$.

In our study of strong lensing of gravitational waves  we consider only one image and not the sum of both, since we expect to see a short burst of GWs which comes only from one image. The second image is delayed in time, with typical time delay of the order of a few months (see e.g. ~\cite{Oguri:2018muv}), much longer than the GW burst. Since we are interested in magnification, we shall compute the cross section for the stronger image. This point has been neglected in the previous literature: usually in eq.\,\eqref{e:cross} $y^{-1}=\mu_1-1$ is replaced by $y^{-1}=\mu/2=(\mu_1+\mu_2)/2$ which is the correct expression for a static situation where both images are seen together. For strong amplification, $\mu_1\sim\mu_2\gg 1$ this difference reduces the cross section by a factor $4$.

For $y<1$, i.e. in the situation where we have two images,  we can rewrite eq.\,(\ref{e:cross}) also in terms  of the time delay between the two images, using the relation (\ref{Deltat}). Explicitly one finds  the following expression for the cross section for time delay smaller than $\Delta t$ 
\be\label{e:cross2}
\sigma(\Delta t,z_{\ell},z_s,\si_v)=\frac{1}{64\pi}\left(\frac{d_s}{d_{\ell}}\right)^2\sigma_v^{-4}\frac{1}{(1+z_{\ell})^2}\Delta t^2\,,
\ee
which is a function of lens and source redshifts and of the velocity dispersion of the lens $\sigma_v$.

\subsection{Optical depth and lensing probability}

To compute the optical depth to the source for magnification of at least $\mu$, $\tau(\mu,z_s)$ we  need to know the physical density $n(\si_v,z_{\ell})$ of lenses (galaxies) with a given velocity dispersion $\si_v$ at redshift $z_{\ell}$. 
We define the density function $n(\si_v,z_{\ell})$ such that its integral $\int n(\si_v,z_{\ell})d\si_v$ simply gives the total density of lenses at redshift $z_{\ell}$.  Then (do not mixup the cross section, $\si$ and the velocity dispersion, $\si_v$)
\be
\int \sigma(\mu,z_{\ell},z_s,\si_v)n(\si_v,z_{\ell})d\si_v\,,
\ee
is the  rate  of scatterings leading to a magnification larger than $\mu$ and 
\be\label{e:depth1}
\tau(\mu,z_s) =\int_0^{z_s}dz_{\ell}\frac{dr}{dz_{\ell}}\textcolor{black}{\left(\frac{d_{\ell}}{d_s}\right)^2}\int \sigma(\mu,z_{\ell},z_s,\si_v)n(\si_v,z_{\ell})d\si_v\,,
\ee
is the optical depth for lensing with magnification $\geq\mu$ (of the most strongly magnified image in the case of two images)  for a source at redshift $z_s$.
Here $dr$ is the physical length element at redshift $z_{\ell}$,
\be
\frac{dr}{dz_{\ell}} =  \frac{1}{(1+z_{\ell})H(z_{\ell})}\,.
\ee
\textcolor{black}{Note also that we have rescaled the cross section to the lens redshift, $\si \ra \si(d_\ell/d_s)^2$ since in eq.\,(\ref{e:depth1}) we multiply by the lens density at $z_\ell$.} 
Inserting this and \eqref{e:cross} for the cross section in eq.\,(\ref{e:depth1}) 
we obtain
\be\label{e:depth2}
\tau(\mu,z_s) =\frac{\pi(4\pi)^2}{(\mu-1)^2}\int_0^{z_s}\!\!\!dz\frac{\chi^2(z,z_s)\chi^2(z)}{\chi(z_s)^2(1+z)^3H(z)}\int_0^\infty d\si_v\si_v^4 n(\si_v,z) \,.
\ee
This is the optical depth for magnification \emph{larger than} $\mu$.

Often only the strong lensing case is considered and the magnification from the two images is added to give the total magnification. To do this one has to replace $1/(\mu-1)^2$ by $4/\mu^2$. As already mentioned above, here we cannot do this since in the case of strong magnification and double images we expect a considerable time delay, so that typically we observe only one image at one given time. Here we assume this to be the stronger image. In the strong magnification case, $\mu\gg2$, this difference is roughly a factor $4$, while in the limit $y\ra 1$ where $\mu\ra 2$ and the second image disappears, the two expressions agree. The probability of having an event from redshift $z_s$ magnified more than $\mu$ is\footnote{To understand eq.~(\ref{e:P>mu}) note that $d\tau/dz$ is the scattering rate leading to a magnification larger than $\mu$ (per redshift). Hence the probability for magnification larger than $\mu$ satisfies the differential equation $dP(>\mu,z)/dz = (1-P)d\tau/dz$ with solution (\ref{e:P>mu}). The factor $(1-P)$ is essential to keep the probability normalized also when $\tau$ becomes large.  In the limit of small optical depth, $P(>\mu,z_s)=\tau(\mu,z_s)$.}
\be\label{e:P>mu}
P(>\mu,z_s)=1-\exp(-\tau(\mu,z_s)) = \int_\mu^\infty p(\mu,z_s)d\mu\,,
\ee
where $p(\mu,z_s)$ is the probability density for  magnification by $\mu$,
\be
p(\mu,z_s) = -\frac{d}{d\mu}P(>\mu,z_s)=-\frac{d\tau}{d\mu}\exp(-\tau(\mu,z_s))\,.
\ee

We can repeat similar steps to compute the probability that a source is strongly lensed and two images are produced with a time separation smaller than $\Delta t$. The optical depth is given by 
\be\label{e:depthT}
\tau(\Delta t, z_s) =\int_0^{z_s}\frac{dz_{\ell}}{(1+z_{\ell})H(z_{\ell})}\textcolor{black}{\left(\frac{d_{\ell}}{d_s}\right)^2} \int\Theta(1/2-y) \sigma(\Delta t,z_{\ell},z_s,\si_v)n(\si_v,z_{\ell})d\si_v\,,
\ee
where the Heaviside function, $\Theta$, has been introduced in order to request that the luminosity of the weaker image is not less than $1/3$ that of the stronger image. This is an arbitrary choice, we just need to assure that both images are above threshold.  Replacing the factor $1/3$ by some arbitrary factor $q$ corresponds to replacing the bound $1/2$ in the Heaviside function by $(1-q)/(1+q)$. Note that here $y(z_s,z_{\ell},\De t,\si_v)$ has to be considered as function of the lens and source redshift for fixed $\De t$ as given in eq.~\eqref{Deltat}.  In a realistic study of time delay one would simply replace our Heaviside function by a minimum signal-to-noise for the weaker image. 
 We can also compute the probability density for a time delay of $\Delta t$ as 
\be
p(\Delta t, z_s) =-\frac{d\tau}{d\Delta t}\exp(-\tau(\Delta t, z_s))\,. 
\ee

The probability densities and probabilities defined in this section depend only on the lens model and lens distribution but not on the distribution of sources. We also note that they are detector independent functions  which depend on the \emph{cosmological} source redshift, which is a quantity that can not be accessed  observationally (without assuming the existence of an optical counterpart) as it is degenerate with the (unknown) magnification. The physically interesting quantity is rather the probability that a source that we \emph{observe} from a given \emph{observed} distance has been magnified by a given amount. We will introduce this quantity in the next section and we will see that it depends on both the distribution of sources and the specifications of a given instrument.

\section{A statistical approach}\label{e:stat}

In this section we answer the following questions: what is the probability that a source from a given \emph{observed} distance has been magnified by a given amount? What is the average magnification and what is its  variance? We derive  expressions for the probability that a given source seen from a given distance has magnification $\mu$,  as a function of the observed source distance. We stress that this is the physically interesting quantity, as we do not have direct access to the information on the intrinsic mass and the cosmological redshift of the emitting source. We also illustrate the analogous approach for the study of the distribution of time delays. 

\subsection{Average amplification}

Without magnification the flux received on Earth, $F$, from a given source with intrinsic luminosity $L$ and cosmological redshift $z_s$ is
\be
F(L,z_s)= \frac{L}{4\pi D(z_s)^2} \,, \qquad L(F,z_s)= 4\pi D(z_s)^2F \,,\qquad D(z_s)=  (1+z_s)\chi(z_s) \,.
\ee
Here $D(z_s)$ is the luminosity distance to redshift $z_s$. Note that even though the wording here is  borrowed from optical astronomy, in a fixed frequency band, the \emph{luminosity} is simply proportional to the square of the strain and hence can be replaced by the strain as we shall do below.  
A realistic GW detector has a limiting strain sensitivity, usually cast as an SNR limit, $\rho_{\lim}$. A source is detected if $\rho\geq\rho_{\lim}$, where the SNR $\rho$ is defined by~\cite{Finn:1992xs}
\be
\rho^2 = 4\int \frac{|h(f)|^2}{S_n(f)}df  =\int \rho^2(f)df\,.
\label{eq:SNR_def}
\ee
Here $S_n(f)$ is the noise of the detector at frequency $f$ 
(see~\cite{Maggiore:1900zz} for more details).
We want to translate a limiting SNR $\rho_{\lim}$ into a limiting flux,  $F_{\lim}$.
The flux per frequency band, $F(f)$ is related to the energy density per frequency band, $\ep(f)$, of the gravitational wave by 
\be\label{FS}
F(f)=c\ep(f) = \frac{(2\pi)^2c}{16\pi G} f^2|h(f)|^2 =\frac{\pi }{16G} f^2S_n(f) \rho^2(f)\,,
\ee
where we used the standard definition of energy density, see e.g. \cite{Maggiore:1900zz},  and for the last equality we used (\ref{FS})  and set the speed of light to one $c\equiv 1$ as in the rest of this paper. We observe that at fixed noise level and in a small frequency band, the limiting flux is proportional to the square of the limiting SNR,  i.e. $F_{\lim}\propto \rho^2_{\lim}$. For fixed source luminosity, the limiting flux scales like $F_{\lim}\propto\mu$, hence the limiting SNR scales as $\rho_{\lim}\propto\sqrt{\mu}$.

 A flux limit  identifies a region in the plane $(L, D(z_s))$ defined by  $F(L, D(z_s))>F_{\lim}$, where $D_s\equiv D(z_s)=(1+z_s)\chi(z_s)=(1+z_s)^2d_s$  denotes the luminosity distance of the source. Given a source distribution with number density per bin of cosmological redshift and intrinsic luminosity $N(L,z_s)$, the number of sources detectable in  a redshift bin of size $dz_s$ around redshift  $z_s$ without considering amplification is
\be\label{e:Ilim}
d\NN(F_{\lim},z_s)= dz_s\int_{L(F_{\lim},z_s)}^\infty N(L,z_s)dL\,.
\ee
Equivalently, the number of sources that we can detect with an emitted luminosity $L\pm dL/2$ is
\be
d\NN(F_{\lim},L)=dL\int_0^{D_s(F_{\lim}, L)} N(L,z_s)\frac{dz_s}{dD_s}dD_s\,,
\ee
where $D_s(F_{\lim}, L)$ is defined by $F(L,D_s)= F_{\lim}$. 

If an event is magnified by $\mu>1$ we see it even if without magnification its flux would have been $F_{\lim}/\mu< F_{\lim}$. It follows that in the presence of magnification, the lower limit of the  integral \eqref{e:Ilim} can be correspondingly reduced.
The total number of objects which we expect to see from redshift $z_s\pm dz_s/2$ including magnification is
\be\label{Nobs1}
d\mathcal{N}_{\rm obs}(z_s,F_{\lim}) = dz_s\int_1^\infty \!\!d\mu p(\mu,z_s)\int_{L_{\lim}(z_s)/\mu}^\infty N(L,z_s)dL \,,
\ee
where $L_{\lim}(z_s)\equiv L(F_{\lim},z_s)$. 
The quantity which is integrated over $\mu$ is the number of sources that we see from  redshift  $z_s$ in the presence of magnification $\mu$, let us denote it by $d\mathcal{N}(\mu, z_s)$ (here we suppress the argument $F_{\lim}$), i.e. 
\be\label{ForPlot1}
d\mathcal{N}(\mu, z_s)\equiv dz_s\int_{L_{\lim}(z_s)/\mu}^\infty N(L,z_s)dL \,.
\ee
Correspondingly, the mean amplification of a source at redshift $z_s$ is
\be\label{ampl1}
\langle\mu\rangle(z_s) = \frac{\int_1^\infty \!\!d\mu\, \mu \,p(\mu,z_s)d\mathcal{N}(\mu, z_s)}{\int_1^\infty \!\!d\mu\, p(\mu,z_s)d\mathcal{N}(\mu, z_s)} \,,
\ee
which of course depends on the threshold in flux $F_{\text{lim}}$, i.e. on the detector sensitivity. Note that (\ref{ampl1}) is the first moment of the probability distribution of magnification (as function of the cosmological redshift) 
\be\label{PDistr}
\mathcal{P}_z(\mu,z_s)\equiv \mathcal{C}\, p(\mu,z_s)\frac{d\mathcal{N}(\mu, z_s)}{dz_s}\,,
\ee
where $\mathcal{C}$ is a  normalization constant.  
For sufficiently high $\mu$, $p(\mu,z_s) \simeq - d\tau/d\mu\propto \mu^{-3}$, while the number of sources which we see with magnification $\mu$ or more from a redshift bin around $z_s$, defined in eq.\,(\ref{ForPlot1}), becomes independent of $\mu$  for large values of $\mu$. Indeed, in the presence of very high magnifications we simply see all sources that are present.\footnote{This last statement is general and does not depend on the details of the distribution of sources. It follows from the physical assumption that there exist a minimum value for the intrinsic luminosity emitted, $L_{\min}$, and once $\mu$ is such that $L_{\lim}/\mu<L_{\min}$ we see all sources and the luminosity integral no longer depends on $\mu$.} The probability of having magnification bigger than $\mu$ for a given source redshift is given by the integral of the probability distribution of magnification, eq.\,(\ref{PDistr}),  from a minimum value $\mu$. We denote this quantity as  $\mathcal{P}_z(>\mu)$. For large values of magnification $\mathcal{P}_z(>\mu)\propto \mu^{-2}$ and the variance of the distribution, defined by $\text{Var}(\mu)\equiv \langle \mu^2\rangle-\langle \mu\rangle^2$ is log-divergent. We stress that this is a general property of the probability distribution of magnification in strong lensing. It does not depend on the details of the distribution of sources, nor on the details of the lens distribution. Moreover, it applies also to the case of transient electromagnetic sources lensed by galaxies: the first moment of the magnification (the mean) is well defined, but the variance is log-divergent. Of course this divergence is not physical. When the magnification becomes very large, $y$ is very small and the geometric optics treatment adopted here breaks down. Nevertheless,  this indicates that the distribution has a significant tail with large magnification. Even though, as we shall see, the probability for a magnification of e.g. 10 is very small, it is not exponentially small as for a Gaussian distribution but it only decays as a mild power law.

The functions determined here, however, are not observable as we typically do not know the cosmological redshift of a GW event. We usually infer it by assuming that the observed 
luminosity distance is the one of the background universe without magnification. This is of course not true when the magnification is relevant. We therefore 
  rewrite (\ref{Nobs1}) and (\ref{ampl1}) as functions of the observed luminosity distance  which is extracted from observations. This is given by
\be\label{dobsdef}
D_{\text{obs}}(z_s,\mu)\equiv\frac{D_s(z_s)}{\sqrt{\mu}}\,.
\ee
In eq.\,(\ref{ampl1}) we also rewrite $p(\mu, z_s)=p(\mu, z_s(D_{\text{obs}}, \mu))$, i.e.  as the probability that a source that appears at a distance $D_{\text{obs}}$ has been amplified by $\mu$. Analogously, 
$dN(L, z_s)=dN(L, z_s(D_{\text{obs}}, \mu))$ is the number of sources emitting  luminosity $L$ which appear at a luminosity distance $D_{\text{obs}}$ associated to redshift $z_s$ and magnification $\mu$ via (\ref{dobsdef}), per unit redshift and intensity. 
The number of sources that we can see from an observed luminosity distance $D_{\text{obs}}\pm dD_{\text{obs}}/2$ is obtained from (\ref{Nobs1}) rewriting all quantities as functions of the observed distance and replacing the redshift bin $dz_s$  by $(dD_{\text{obs}}/dz_s)^{-1}dD_{\text{obs}}=\sqrt{\mu}(dD_{s}/dz_s)^{-1}dD_{\text{obs}}$. This gives 
\bea \label{e:Nobs}
d\NN_{\rm obs}(D_{\text{obs}},F_{\lim}) 
&=&d D_{\text{obs}} \int_1^\infty \!\!d\mu  p(\mu, z_s(D_{\text{obs}}, \mu))\frac{\sqrt{\mu}}{D'_s(z_s(D_{\text{obs}}, \mu))}\int_{L_{\lim}/\mu}^\infty N(L, z_s(D_{\text{obs}}, \mu)) \,dL \,\nonumber,\\
\eea
where  $D_{s}'(z)\equiv dD_s/dz_s$. 
In these expressions $z_s(\mu, D_{\text{obs}})\equiv z_s(D_s=\sqrt{\mu}D_{\text{obs}})$ is the cosmological redshift that corresponds to the observed distance $D_{\text{obs}}=D_s(z_s)/\sqrt{\mu}$. Not to be confused with the redshift \emph{inferred} by an observer who assumes that $D_{\text{obs}}$ is the luminosity distance of the background universe (without magnification). This latter, which we denote by $z_{\text{obs}}$,  is given by $D_s(z_{\text{obs}})=D_{\text{obs}}$.

The quantity integrated in  (\ref{e:Nobs}) over $\mu$ is the number of sources which we see  from a luminosity distance $D_{\text{obs}}\pm dD_{\text{obs}}/2$, with magnification $\mu$. For future use, we define this quantity as  
\be\label{ForPlot2}
d\mathcal{N}(\mu, D_{\text{obs}})\equiv dD_{\text{obs}} \frac{\sqrt{\mu}}{D'_s(z_s(D_{\text{obs}}, \mu))}\int_{L_{\lim}/\mu}^\infty N(L, z_s(D_{\text{obs}}, \mu)) \,dL  \,,
\ee
where it is understood that $L_{\text{lim}}\equiv L(F_{\text{lim}}, D_{\text{obs}})~=4\pi  D_{\text{obs}}^2F_{\text{lim}}$, which, at fixed $D_{\text{obs}}$, is independent of the magnification. 
Note also that like $d\mathcal{N}(\mu, z_s)$, the quantity $d\mathcal{N}(\mu, D_{\text{obs}})$ is independent of the optical depth. It is the (differential) number of sources which are visible with observed luminosity distance $D_{\text{obs}}$ with magnification $\mu$.  
The convolution of this quantity  with the probability density of magnification gives the total number of events observed from a  given observed luminosity distance, eq.\,(\ref{e:Nobs}).
The average amplification as a function of the observed luminosity distance  is therefore
\be\label{ampl2}
\langle\mu\rangle(D_{\text{obs}}) = \frac{\int_1^\infty \!\!d\mu\, \mu \, p(\mu, D_{\text{obs}})d\mathcal{N}(\mu, D_{\text{obs}})/dD_{\text{obs}} }{\int_1^\infty \!\!d\mu\, p(\mu, D_{\text{obs}})d\mathcal{N}(\mu, D_{\text{obs}})/dD_{\text{obs}}} \,,
\ee
which of course depends on the threshold in flux $F_{\text{lim}}$, i.e. on the detector sensitivity. In full analogy with eq.\,(\ref{PDistr}), we  introduce the probability distribution of magnification as function of the observed luminosity distance of the source 
\be\label{PDistr2}
\mathcal{P}_{\text{obs}}(\mu, D_{\text{obs}})\equiv \mathcal{C}\, p(\mu, D_{\text{obs}})\frac{d\mathcal{N}(\mu, D_{\text{obs}})}{dD_{\text{obs}}}\,, 
\ee
where $\mathcal{C}$ is a constant of normalization which is given by the denominator of (\ref{ampl2}). 

From eq.\,(\ref{ForPlot2}), it follows that $d\mathcal{N}(\mu, D_{\text{obs}})/dD_{\text{obs}}\propto \sqrt{\mu}$ for $\mu\gg1$ as both the lower integration bound and $D'_s$ tend to constant asymptotic values for large magnification $\mu$. Note that $L_{\text{lim}} =4\pi  D_{\text{obs}}^2F_{\text{lim}}$ is independent of $\mu$, hence $L_{\text{lim}}/\mu$ tends to zero. The probability density $p(\mu,D_{\text{obs}})$ in \eqref{PDistr2} depends on $\mu$ also through the observed luminosity as we have defined  $p(\mu, D_{\text{obs}})\equiv p(\mu, z_s(\sqrt{\mu}D_{\text{obs}}))$. Recalling that $p(\mu, z_s)\propto \mu^{-3}$, it follows that $p(\mu, D_{\text{obs}})$ will have a milder decay with magnification. Indeed, for a fixed $D=D_{\text{obs}}$,  and $\mu>1$, $p(\mu, D_{\text{obs}})\equiv p(\mu, z_s(\sqrt{\mu}D_{\text{obs}}))$ receives contributions also from redshifts higher than $z_s(D_{\text{obs}})$ and therefore the lensing probability is higher.  The exact scaling of $p(\mu, D_{\text{obs}})$ with $\mu$ depends on the details of the lens distribution.

We anticipate here some results that we will obtain in section \ref{catalogues} when considering a realistic model for the galaxy distribution and evolution.  For a realistic distribution with comoving number density of lenses which reproduces observations (or hydrodynamical simulations calibrated to observations), we find that $p(\mu, D_{\text{obs}})\propto \mu^{-\alpha}$. The value of $\alpha$ increases from 2 to 3 as we increase the observed luminosity distance from $\sim 300$Mpc to $5000$Mpc where it saturates  to the asymptotic value $\alpha=3$. It follows that the probability distribution (\ref{PDistr2}) scales as $\mathcal{P}_{\text{obs}}(\mu)\propto \mu^{-\alpha+1/2}$ and its integral from $\mu$ to infinity behaves as  $\mathcal{P}_{\text{obs}}(>\mu)\propto \mu^{-\alpha+3/2}$ for large values of $\mu$. It follows that, for any fixed value of $D_{\text{obs}}$ the variance of the distribution diverges as $\mu^{-\alpha+7/2}$. We stress that this is a general property of the probability distribution of magnification (for different observed redshift), independent of the details of the distribution of sources. Moreover, it applies also to the case of strongly lensed sources of electromagnetic radiation. Also in that case, the variance of the distribution of magnification (for different observed redshift) is divergent. We observe that for realistic models of lens distribution also the mean of (\ref{PDistr2}) diverges for small observed distances. 
The fact that the variance of the distribution diverges is due to the presence of extremely highly magnified sources, i.e. sources aligned along the lens-observer direction for which the magnification $\mu\rightarrow\infty$, see eq.\,(\ref{e:muy}). This is also at the origin of the divergence of the mean of the magnification at small observed distances. Physically, the magnification tends to a finite value and, as mentioned above,  the presence of the divergence is due to the geometric optics approximation,  which breaks down in the vicinity of caustics, where wave effects need to be taken into account. The fact that the mean of the magnification diverges only for small observed distances is due to the fact that, if a given source is very highly magnified, it appears at a small distance.  We will get back to this point in Sec.\,\ref{s:num} and \ref{s:con}. 

\subsection{Average time delay}

The average time delay can not be obtained in the same way. The problem 
is that when computing the analog of eq. (\ref{ampl2}) for the average time delay, both $z_s(D_{\text{obs}},\mu)$ and the lower bound of the integral over the source luminosity depend on the magnification, while the probability density $p(\Delta t,z_s)$ has contributions from different magnifications. On the other hand, a fixed magnification $\mu$ can lead to different time delays depending on the velocity dispersion $\si_v$ of the lensing galaxy and on its redshift, $z_{\ell}$.  In order to have a double image which can lead to time delay, the magnification of the stronger image is $\mu_1=1+1/y>2$ since $y<1$ is required.

It is convenient in this case to explicitly  keep the dependence on $(z_{\ell}, z_s, \sigma_v, \Delta t)$ and integrate over these only at the very end when computing the average $\Delta t$. We then have for the optical depth, using (\ref{e:cross2})
\begin{align}\label{e:depthT}
\tau(\Delta t, z_s, z_{\ell}, \sigma_v) &=\frac{1}{(1+z_{\ell})H(z_{\ell})}\left(\frac{d_{\ell}}{d_s}\right)^2 \sigma(\Delta t,z_{\ell},z_s,\si_v)n(\si_v,z_{\ell})\nn\\
&=\frac{1}{64\pi(1+z_{\ell})^3H(z_{\ell})}\sigma_v^{-4}n(\si_v,z_{\ell})\,.
\end{align}
We can compute the probability density as 
\be
p(\Delta t, z_s, z_{\ell}, \sigma_v)  =-\frac{d\tau}{d\Delta t}\exp(-\tau(\Delta t, z_s, z_{\ell}, \sigma_v) )\,.
\ee
The average time delay for a source observed at luminosity distance $D_{\text{obs}}$ is given by
\be\label{amplT}
\langle\Delta t\rangle(D_{\text{obs}}) = \frac{\int_0^\infty \!\!d\Delta t \int dz_{\ell}\int d\sigma_v\,\Delta t \, p(\Delta t, D_{\text{obs}}, z_{\ell}, \sigma_v)\left(dz_s/dD_{\text{obs}}\right)\int_{ L_{\lim}/\mu_2}^\infty N(L, D_{\text{obs}}) \,dL}{d\mathcal{N}_{\text{obs}}(D_{\text{obs}},F_{\lim})} \,, 
\ee
where we are considering the magnification of the fainter image giving the lowest integration bound over intensity to make sure that both images are seen 
\be
\mu_2(\Delta t, D_{\text{obs}}, z_{\ell}, \sigma_v)=\frac{\beta_t(\si_v^2,z_\ell,z_s)}{\Delta t}-1\,, 
\ee
where the function $\beta_t$ is defined in eq.\,(\ref{Deltat}).

\section{A simple model for semi-analytical results}\label{s:ana}

Let us first discuss a simplified analytical model to have some intuition of the size of the mean magnification as a function of redshift. The advantage of an analytic description is that it allows us to analyze the dependence of the results on the various astrophysical ingredients entering the modelling. We need two basic ingredients: a model for the lens distribution in order to derive the probability density of lensing, and a model for the distribution of sources as a function of redshift and intensity. 

\subsection{Analytic description of lens distribution}\label{RuthMod}

If there would be no evolution of lenses we would expect the physical number density of galaxies to scale with redshift as $n(\si_v,z)\propto (1+z)^3$, hence  
\be\label{e:evol-ass}
\int_0^\infty d\si_v\si_v^4 n(\si_v,z)  = (1+z)^3N\langle\si_v^4\rangle \,,
\ee
where $N$ is the present galaxy density,
\be\label{e:Nz0}
N= \int_0^\infty d\si_v n(\si_v,z=0) \,,
\ee
 and $\langle\si_v^4\rangle$ a mean of the velocity dispersion to power 4. Let us now assume that in the considered redshift range it is legitimate to neglect evolution, hence the physical density just scales with the expansion of the Universe. We postulate this law of evolution not because it is very realistic, but because it allows us to integrate the optical depth analytically. However, as we shall see  in the next section, for low redshifts to which present gravitational wave experiments are sensitive, this model is actually quite good. 
 
 With the assumption \eqref{RuthMod}, the factor $1/(1+z)^3$ in \eqref{e:depth2} cancels and using $ \chi(z,z_s)'=-H(z)^{-1}$ we  obtain 
\be\label{e:tau-ana}
\tau(\mu,z_s) =\frac{\pi(4\pi)^2\langle\si_v^4\rangle N}{\chi_s^2}(\mu-1)^2\int_0^{\chi_s}\!\!\!d\chi(\chi_s-\chi)^2\chi^2
= \frac{\pi(4\pi)^2\langle\si_v^4\rangle N\chi_s^3}{30(\mu-1)^2}\,.
\ee
Crude estimates for $N$ and $\si_v$ are 
\be
N = 10^{9}H_0^3 \,, \qquad \langle\si_v^4\rangle =(150\rm{km/s})^4 \simeq 5\times10^{-14}\,.
\ee
As we have set $c=1$ we have $H_0 =h/(3000\text{Mpc})$. 
With this we obtain for the optical depth 
\be\label{e:simp1}
\tau(\mu,z_s) \simeq \frac{0.001}{(\mu-1)^2}(H_0\chi(z_s))^3 \,.
\ee
This analytic model gives a result for optical depth in very good agreement with the one obtained in section \ref{Ill} considering a more realistic distribution of lenses, which evolve with redshift (fractional deviations of a few percent). We will get back to the effects of galaxy evolution on the probability of lensing  in section \ref{Ill}, see in particular Fig.\,\ref{Pmu}. 
For our simple model we find for the probability density of magnification as a function of the cosmological source redshift 
\be
p(\mu,z_s) =
\frac{2p_1(z_s)}{(\mu-1)^3}\exp\left(-\frac{p_1(z_s)}{(\mu-1)^2}\right) \,, \quad
\mbox{where } \qquad
p_1(z_s)=0.001 (H_0\chi(z_s))^3\,.
\ee
This simplified analytical model for the lens distribution will allow us in the remaining of this section to compute analytically the average amplification of a source at a given redshift $z_s$ and to explore how the result depends on the astrophysical model for the source population.

\subsection{Compact binary population: analytical description}\label{an:sources}

In this section we work out an analytic model to describe the number density of sources as a function of the emitted gravitational-wave strain amplitude (or the SNR for a given detector configuration) and as a function of redshift. As a first step we review the derivation of the SNR of a binary system of compact objects at fixed redshift as a function of the component masses and of the noise power spectral density of the instrument(s).  
We introduce the SNR per unit frequency $\rho$ for a binary that merges at redshift $z_s$ and is observed in a {narrow} frequency band around $f$ as~\cite{Finn:1992xs}
\begin{equation}\label{Finn0}
 \rho^2(f, z_s)=4\frac{|h(f, z_s)|^2}{S_n(f)}\propto \frac{F(f, z_s)}{S_n(f)}= \frac{L(f, z_s)}{4\pi D(z_s)^2S_n(f)}\,,
\end{equation}
where $S_n(f)$ is the noise power spectrum of a given detector network, with dimension of Hz$^{-1}$ and $h(f, z_s)$ is the Fourier transform of the strain, with dimensions \text{Hz}$^{-1}$. The total SNR in the frequency band of a given detector is simply the integral of eq.\,(\ref{Finn0}) over that band. In eq.\,(\ref{Finn0}) we have defined the received flux per unit of  frequency as $F(f, z_s)$ and the related luminosity per unit of (observed) frequency as $L(f, z_s)$. Hence  there is a one-to-one mapping between flux and SNR or luminosity and SNR. For a single binary system the SNR from the inspiral phase integrated over observed frequency is given by~\cite{Finn:1992xs} (neglecting the merger and ringdown phases and writing the quantities on which the SNR depends explicitly)
\be\label{Finn}
\rho^2(z_s, m, \mathcal{M})= \frac{5}{96\pi^{4/3}} \frac{\Theta^2}{D^2(z_s)} (G \mathcal{M}_z)^{5/3} f_{7/3}(z_s, m)\,,
\ee
where $\mathcal{M}_z=\mathcal{M}(1+z_s)$ is the redshifted  chirp mass of the system and $m=m_1+m_2$ is the total mass and the integral over frequencies can be approximated by
\be\label{f73}
f_{7/3}(z_s, m)\equiv \int_0^{f_{\text{ISCO}}/(1+z_s)} df\left[f^{7/3} S_n(f)\right]^{-1}\,,
\ee
with dimension $[f_{7/3}]=[\text{Hz}]^{-1/3}$  while $\Theta^2$ (with $0\leq\Theta^2\leq16$) is a geometrical factor that depends on the inclination of the binary and on the antenna pattern of the detector. Averaging over all angles one finds $\overline{\Theta^2}=64/25$. We do not need this explicit expression in what follows and the interested reader is referred to~\cite{Finn:1992xs} for details. In eq.\,(\ref{f73}) the upper integration bound is given by the (redshifted)  frequency corresponding to the innermost stable circular orbit ($\text{ISCO}$) of the system, i.e. the frequency at which we consider the inspiraling phase of the system to end in our approximation. It is defined as \cite{Maggiore:1900zz}
\be\label{ISCO}
f_{\text{ISCO}}\equiv\frac{1}{6\sqrt{6}(2\pi)}\frac{c^3}{Gm}\simeq  2.2. \text{kHz}\left(\frac{M_{\odot}}{m}\right)\,,
\ee
where $m$ is the total mass of the binary system. This frequency is not the frequency where the signal is maximal, but it is the frequency beyond which we can no longer trust the quadrupole formula used for the SNR estimate in eq.~\eqref{Finn}.
It is therefore a conservative estimate of the SNR which has contributions from higher frequencies, e.g. the coalescence frequency is typically about $3f_{\text{ISCO}}$. But we follow here the standard procedure as detailed in Ref.~\cite{Finn:1992xs}. 
We rewrite the parametrization (\ref{Finn}) in a way which will be useful for what follows,
\be\label{Giulia}
\rho^2(z_s, \mathcal{M}, m)=7.7 \cdot 10^{-39} \Theta^2 \left(\frac{\mathcal{M}}{M_{\odot}}\right)^{5/3}\left(\frac{\text{Mpc}}{D(z_s)}\right)^2\frac{f_{7/3}(z_s, m)}{\text{Hz}^{-1/3}}(1+z_s)^{5/3}\,.
\ee
\begin{figure}[htp!]\center
\includegraphics[width=1.05\textwidth]{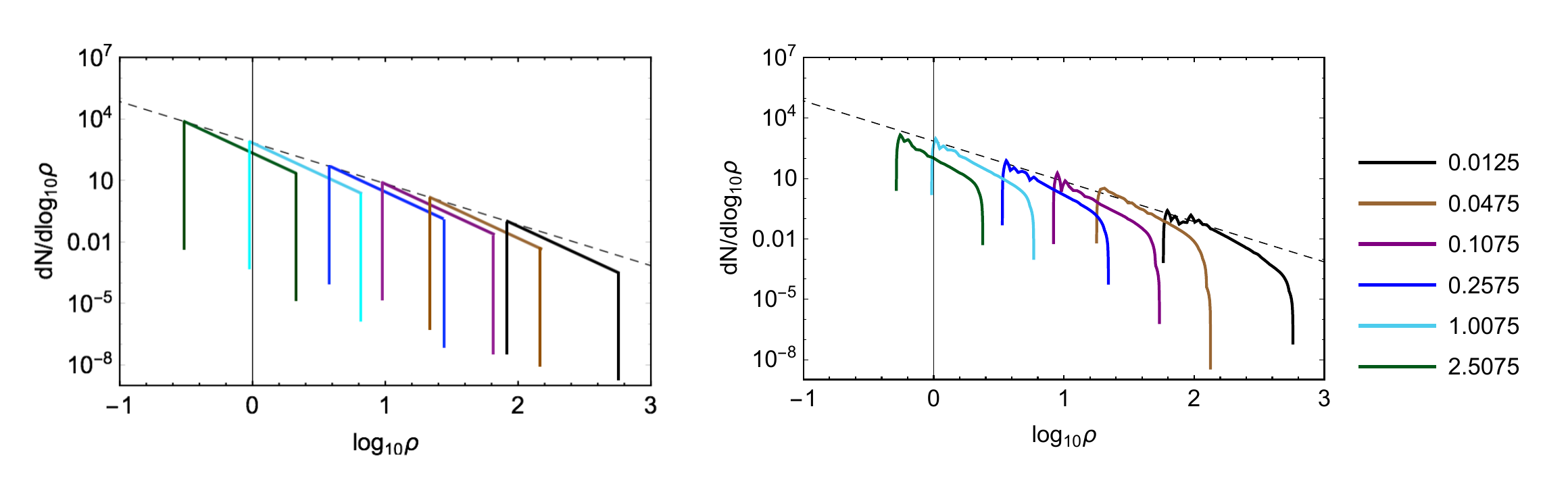} 
	\caption{\label{Fin}Comparison between the analytic model for source distribution presented in section \ref{an:sources} and the numerical simulations of section \ref{catalogue}, for different redshifts $0.0125\leq z \leq 2.5075$. }
\end{figure}
Let us understand the redshift dependence: from eq.\,(\ref{f73}) we see that if $f_{\text{ISCO}}/(1+z_s)> f_{\text{max}}$  then $f_{7/3}(z_s)$ does not depend on redshift. Here with $f_{\text{max}}$ we denote the maximum frequency a given detector is sensitive to (i.e. the frequency above which the noise becomes very large) and it is fixed by the form of the spectral noise in  eq.\,(\ref{f73}). This is determined by the condition
\be\label{ISCOCondition}
(1+z_s)<\frac{f_{\text{ISCO}}}{f_{\text{max}}}\simeq\left(\frac{2.2.\,\text{kHz}}{f_{\text{max}}}\right)\left(\frac{M_{\odot}}{m}\right)\,. 
\ee
If this condition is verified for a given source, then (\ref{f73}) is redshift independent. 

From a comparison with numerical models for source formation and evolution (see section \ref{s:num}), we find that the maximum SNR that we receive from a given redshift $z_s$ can be effectively written as 
\be\label{max}
\rho_{\text{max}}(z_s)= \mathcal{A} \left(\frac{\mathcal{M}_{\text{max}}}{M_{\odot}}\right)^{5/6}\frac{\text{Mpc}}{D(z_s)}(1+z_s)^{5/6}\bar{f}_{7/3}(z_s)^{1/2}\,,
\ee
and similarly for the minimum value
\be\label{min}
\rho_{\text{min}}(z_s)= \mathcal{A} \left(\frac{\mathcal{M}_{\text{min}}}{M_{\odot}}\right)^{5/6}\frac{\text{Mpc}}{D(z_s)}(1+z_s)^{5/6}\bar{f}_{7/3}(z_s)^{1/2}\,,
\ee
where $\bar{f}_{7/3}(z_s)^{1/2}$ is obtained using the average ISCO frequency, rescaled with redshift,  as upper cutoff in the integral~\eqref{f73}, i.e. the ISCO frequency for the average total mass $m$. We have introduced a constant $\mathcal{A}=1.4\cdot 10^{-19}$ which corresponds to the square root of the numerical factor in eq.\,(\ref{Giulia}) multiplied by the square root of $\bar{\Theta}^2=64/25$.  We denoted as $\mathcal{M}_{\text{min}}$ and $\mathcal{M}_{\text{max}}$ the minimum and maximum value of the chirp mass (for a given mass distribution). We observe that for redshifts that satisfy the condition (\ref{ISCOCondition}), eqs.\,(\ref{min}) and (\ref{max}) scale with redshift simply as $(1+z_s)^{5/6}D(z_s)^{-1}$.

As a second step, we need a parametrization for the number of sources per unit of SNR as a function of SNR and of redshift. The number density of sources per unit of chirp mass (and as a function of redshift) is given by 
\be\label{NMc}
N(z_s,  \mathcal{M}_c) \propto \mathcal{R}(z_s,  \mathcal{M}_c) \propto p(\mathcal{M}_c)\,,
\ee
where $\mathcal{R}$ is the merger rate for a binary system with chirp mass $\mathcal{M}_c$ at  redshift $z_s$ which in turn is proportional to the distribution of chirp masses (with  a redshift-dependent pre-factor). For simplicity we assume it to be polynomial, $p(\mathcal{M}_c)\propto \mathcal{M}_c^{-\beta_M}$. 
We need now to convert the number density (\ref{NMc}) into a number density per unit of SNR. Neglecting the mass dependence of the ISCO frequency, as we did in (\ref{min}) and (\ref{max}), we can simply invert (\ref{Giulia}) and we get for the distribution of sources as a function of SNR
\be\label{NI3}
N(\rho,z_s)= \left\{\begin{array}{ll} N_{\rho}(z_s)\rho^{-\gamma}  & \mbox{for } \rho_{\text{min}}(z_s)<\rho<\rho_{\text{max}}(z_s)\,,\\
 0  & \mbox{otherwise,}\end{array} \right.
\ee
with $\gamma=(6\beta_M-1)/5$ and $N_{\rho}(z_s)$ is some redshift dependent function. 
For example, the distribution of masses in the astrophysical model discussed in Section~\ref{s:num} is such that the resulting distribution of chirp masses is (independent on redshift) $p(\mathcal{M}_c)\propto \mathcal{M}_c^{-3.2}$. Then $N(\rho, z_s)\propto \rho^{-3.6}$, which is the scaling of the curves in Figs.\,\ref{Fin} and \ref{pop}.\footnote{Note that when comparing these figures to the scaling derived here, one needs to keep in mind that $dN/d\rho=\rho^{-1} dN/d\log\rho$.}

\subsection{Analytic computation of the average magnification}\label{muav}

We have now all the ingredients necessary to compute the average magnification of sources located at a given redshift, for a generic detector network. 
The number of sources which we see from redshift $z_s$ with our limiting SNR $\rho_{\lim}$ when magnified by $\mu$ is
\begin{align}\label{e:Nlim2}
N(\rho_{\lim},z_s,\mu)&= N_{\rho}(z_s)\int_{\alpha_{\lim}}^{\rho_{\text{max}}(z_s)} d\rho\,\rho^{-\gamma}\,,
\end{align}
where $\alpha_{\lim}=\text{max}\left(\rho_{\lim}/\sqrt{\mu}\,, \rho_{\text{min}}(z_s)\right)$. We assume $\rho_{\lim}/\sqrt{\mu}<\rho_{\text{max}}(z_s)$. If this is not the case, $N(\rho_{\lim},z_s,\mu)=0$. 
The average magnification can then be computed analytically as
\be\label{e:mu}
\langle \mu \rangle (\rho_{\text{lim}}, z_s)=\int_1^\infty d\mu \,\mu\, p(\mu,z_s)N(\rho_{\lim},z_s,\mu)\left(\int_1^\infty d\mu \,p(\mu,z_s)N(\rho_{\lim},z_s,\mu)\right)^{-1}\,. 
\ee
The explicit analytic expression of the average amplification as a function of the parameters of the model and its derivation can be found in appendix \ref{appendixMu}. 

Our analytic model for the distribution of sources has four free parameters: ($\gamma$, $\mathcal{M}_{\text{min}}$, $\mathcal{M}_{\text{max}}$) defining the distribution of chirp masses and $\rho_{\text{lim}}$ (or equivalently $S_n$) which gives the minimum SNR for detection, and depends on the detector network chosen. The computation of the average magnification presented in this section relies on the assumption that we can replace the mass dependence of the ISCO frequency appearing in the expression of the SNR (\ref{Giulia}) by an average. From a  comparison with numerical models of section \ref{s:num}, it turns out that this approximation is well justified for low redshift sources ($z_s<2$ when considering the sensitivity curve of the LIGO-Virgo detectors). The comparison is shown in Fig.\,\ref{Fin}. As already explained, the average magnification does not depend on the scaling of the number of events with redshift, i.e. on $N_{\rho}(z_s)$ in eq.\,(\ref{NI3}). To compare with a realistic population model,  in Fig.\,\ref{Fin} we have reconstructed this dependence fitting the numerical model, i.e.  $N_{\rho}(z_s)\propto \rho^{-3}$.  For high redshift sources a simple analytic derivation of the average magnification is not possible, but the relation (\ref{Giulia}) can be inverted numerically to find the chirp mass and the integral  (\ref{e:mu}) can be performed numerically to find the average magnification as a function of redshift.

\subsection{Analytical results for the LIGO detector(s)}

\begin{figure}[htp!!]\center
\includegraphics[width=0.55\textwidth]{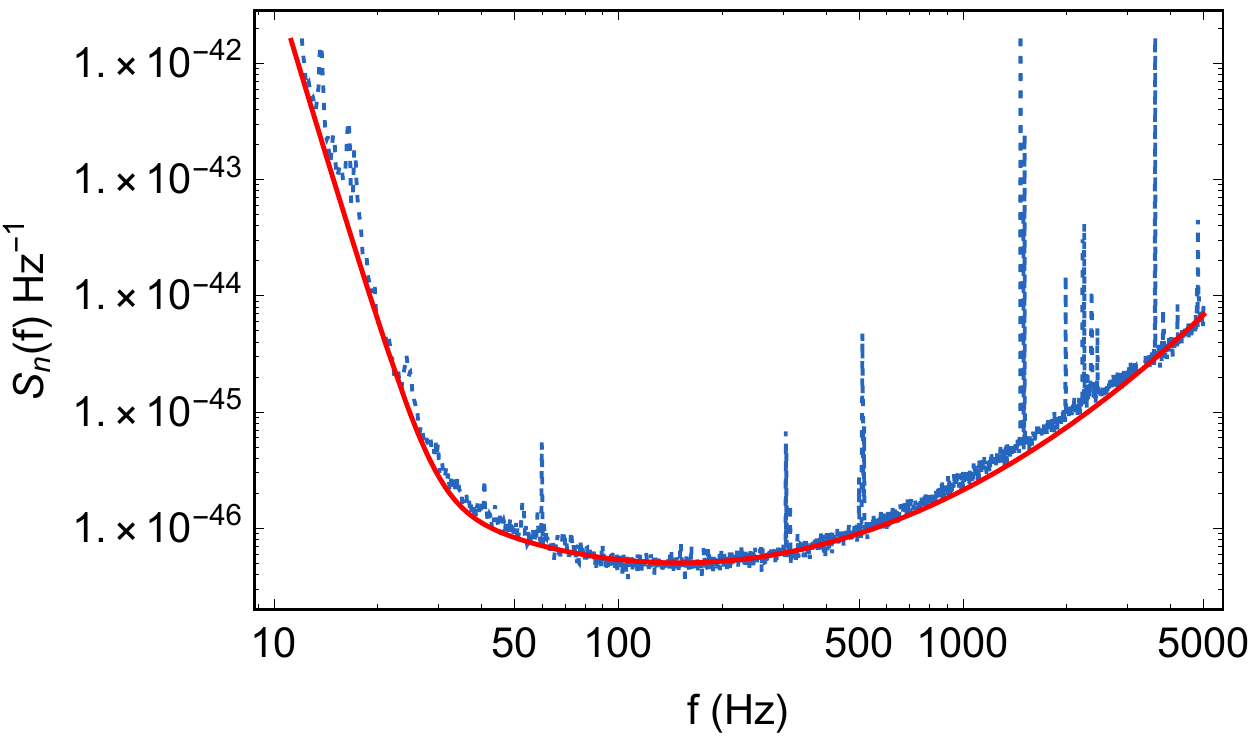} 
	\caption{\label{Livingstone}Spectral noise curve for the LIGO-Livingstone detector \cite{LIGOWeb}. The red curve is the fit in eq.\,(\ref{fit}).}
\end{figure}

 In this section we focus on the case of the LIGO detectors and we compute the average magnification for different astrophysical models for the distribution of sources, corresponding to different choices of the parameters  ($\gamma$, $\mathcal{M}_{\text{min}}$, $\mathcal{M}_{\text{max}}$). For definiteness we consider the noise curve of the LIGO-Livingstone detector. In Fig.\,\ref{Livingstone} we plot the spectral noise of the LIGO-Livingstone detector that we find to be well described by the fitting function $S_{\text{fit}}(f)$ given by
\be\label{fit}
S_{\text{fit}}(f)=\mathcal{A}_s\left[\left(\frac{f}{30 \text{Hz}}\right)^{\alpha_s}+1\right]\exp\left[\beta_s\left(\log\left(\frac{f}{150\text{Hz}}\right)\right)^2 \right]\,,
\ee
with $\log_{10}\mathcal{A}_s=-46.3$, $\alpha_s=-7.8$ and $\beta_s=0.4$.

\begin{figure}[hpt!!]
\includegraphics[width=0.5\textwidth]{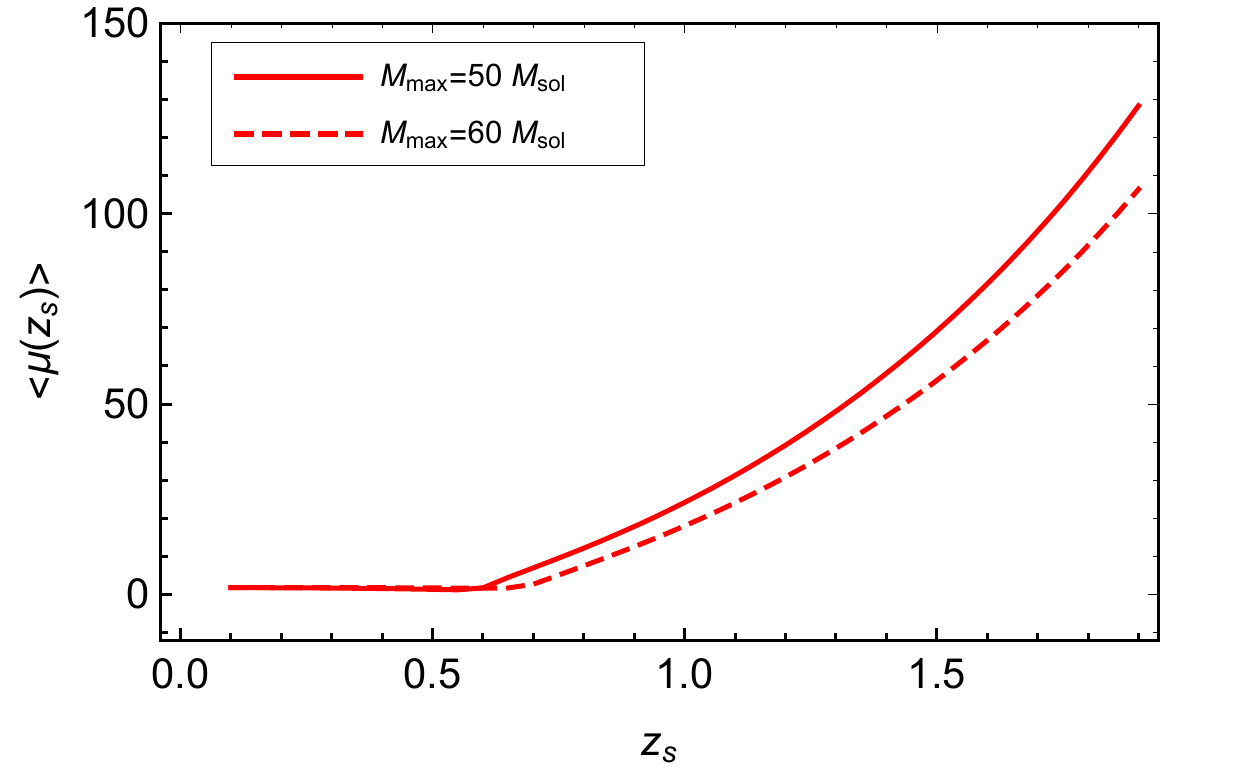} \quad 
\includegraphics[width=0.5\textwidth]{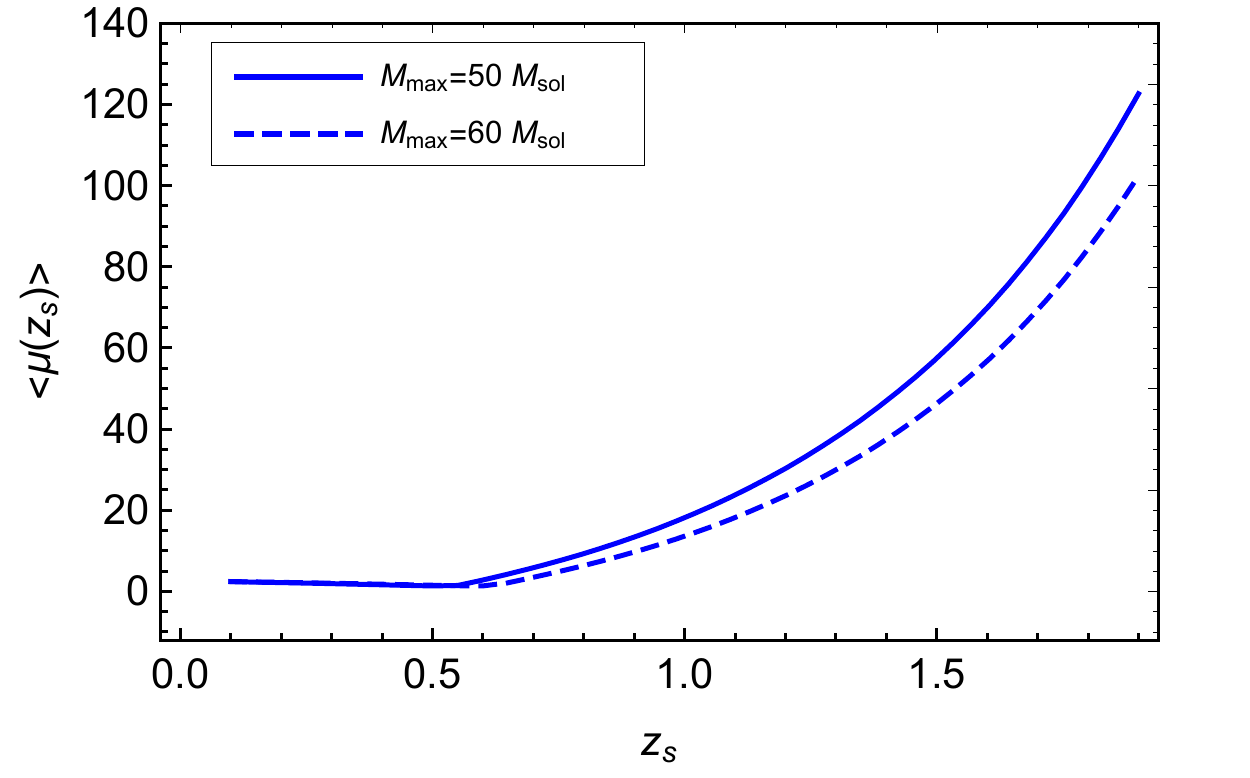} 
	\caption{Left: Average magnification for  $\alpha=2.35$ ($\gamma=3.6$) for two different upper cut-offs in the distribution of the primary mass. Right: Average magnification for  $\alpha=1.5$ ($\gamma=2.3$) for two different upper cut-offs in the distribution of the primary mass.}
	\label{An1}
\end{figure}

We analyze the dependence of the average amplification on the details of the simple astrophysical model presented in Sec.\,\ref{muav}. In particular, we want to explore the dependence on the distribution of the component masses and on the upper cut-off in the distribution.  
We assume that the primary BH mass is distributed as $p(m_1)\propto m_1^{-\alpha}$ 
in the range $[5, X]M_{\odot}$ where both $\alpha$ and $X$  are parameters that we vary. We consider two cases: $\alpha=2.35$ and $\alpha=1.5$. The corresponding normalized distribution of chirp masses is 
polynomial in the chirp mass, i.e. $p(\mathcal{M}_c)\propto \mathcal{M}_c^{-\beta_M}$ with $\beta_M=3.2$ and $\beta_M=2.1$ respectively. As explained in the previous section, for sufficiently small redshift, the scaling of the number density of sources per unit of signal to noise is  given by (\ref{NI3}) with  $\gamma=-1/5(1-6\beta_M)$, hence $\gamma=3.6$ and $2.3$ for the two values of $\alpha$, respectively. In Fig.\,\ref{An1} we show results for the average magnification as a function of redshift for the two models considered and for two different values of  the upper cut-off in the mass distribution.  
As we will see in section \ref{s:av}, the result for the average amplification in Fig.\,\ref{An1} is in very good agreement with the one we find using a realistic model for source formation and evolution, for low redshift sources (see Fig.\,\ref{mu2}, left panel). 

We have considered here the case of the LIGO-Livingston detector, but analogous results can be easily obtained for any other ground based detector (or detector network), just using the noise curve of that detector.  For example, when using the Einstein Telescope (ET) where significantly lower frequencies and lower mass objects can be detected (see Fig.~\ref{Irina} for the noise curve),   we expect that our analytical model can be extrapolated up to higher redshift.  
Indeed, comparing our analytical model with the numerical one, we find (see Section~\ref{s:num}) that it is a reasonably good fit as long as high mass binaries have a higher SNR than low mass ones. For the ET which  has much better sensitivity at 1-10Hz this remains true up to higher redshifts than for LIGO-Virgo.

\FloatBarrier

\section{Numerical results}\label{s:num}

In this section we model the redshift distribution and redshift evolution of lenses (galaxies) using fits to hydrodynamical simulations. We also model the distribution of GW sources using a phenomenological model of the binary BH population. We then compute the distribution of magnification as a function of the cosmological source redshift and of the observed one. We focus on the case of the two LIGO/Virgo detectors. Finally we compute the probability that the GW events observed during O1+O2 observation campaign have been magnified by a given amount.

\subsection{The lens distribution}\label{catalogues}

\begin{figure}[ht!]
\begin{center}
\includegraphics[width=0.45\textwidth]{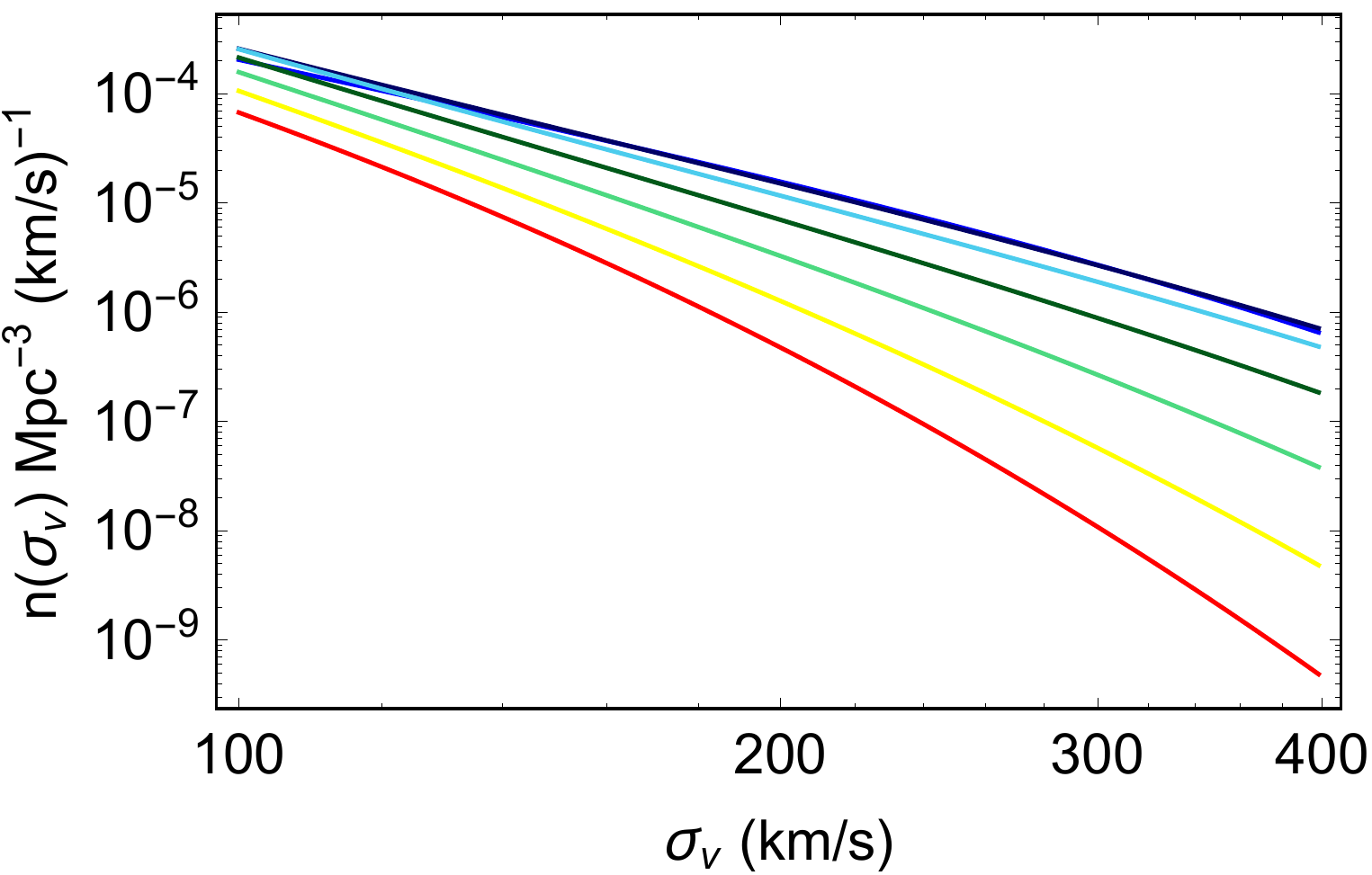}  
\end{center}
\caption{\label{NN} Comoving number density of galaxies as a function of  of velocity dispersion  $\si_v$ for different redshifts. The color code is:  $z=0, 1, 2, 3, 4, 5, 6$ from black to red. The figure is obtained from fits to the Illustris simulation  of \cite{2015MNRAS.454.2770T}.}
\end{figure}

We  model the number density of galaxies (lenses) as a function of the velocity dispersion $\sigma_v$, taking into account the evolution of galaxies with redshift.  For this, we use the results of Torrey et al. \cite{2015MNRAS.454.2770T} based on the Illustris hydrodynamical simulation. We  use the values in Table 6 in the ArXiv version of 
Ref.~\cite{2015MNRAS.454.2770T}, but we do not use the fitting formula given there. The correct fit is~\cite{Paul}
\begin{align}\label{fit2}
\log_{10} N(>\sigma_v, z)=A(z)+\alpha(z)
\left(\log_{10}\sigma_v-\gamma(z) \right)+\beta(z)\left(\log_{10}\sigma_v-\gamma( z)\right)^2 -\left(\sigma_v\times10^{-\gamma(z)}\right)^{1/\ln(10)}\,,
\end{align}
where the numerical value of $\sigma_v$ is to be taken in units of km/sec. We stress that here $\sigma_v$ represents the \emph{redshift independent} velocity dispersion.  The functions $A$, $\al$, $\beta$ and $\gamma$ are modeled as
\begin{align}
A&=a_0+a_1 z+a_2 z^2\,,\\
\alpha&=\alpha_0+\alpha_1 z+\alpha_2 z^2\,,\\
\beta&=\beta_0+\beta_1 z+\beta_2 z^2\,,\\
\gamma&=\gamma_0+\gamma_1 z+\gamma_2 z^2\,.
\end{align}
For completeness we give the parameters $a_0$ to $\gamma_2$ in Appendix \ref{app:Illustris}. 
The central stellar velocity dispersion,  $\sigma_v$ is defined as the three-dimensional standard deviation of the stellar velocity within the stellar half-mass radius. We   define the comoving number density of galaxies as 
 \be
n^{\text{com}}(\sigma_v, z)\equiv - \frac{d N}{d\sigma_v}(>\sigma_v, z)\,.
\ee
The result for different redshifts is shown in Fig. \ref{NN}. 

\subsection{Results for optical depth and probability}\label{Ill}

\begin{figure}[ht!]
\begin{center}
\includegraphics[width=0.45\textwidth]{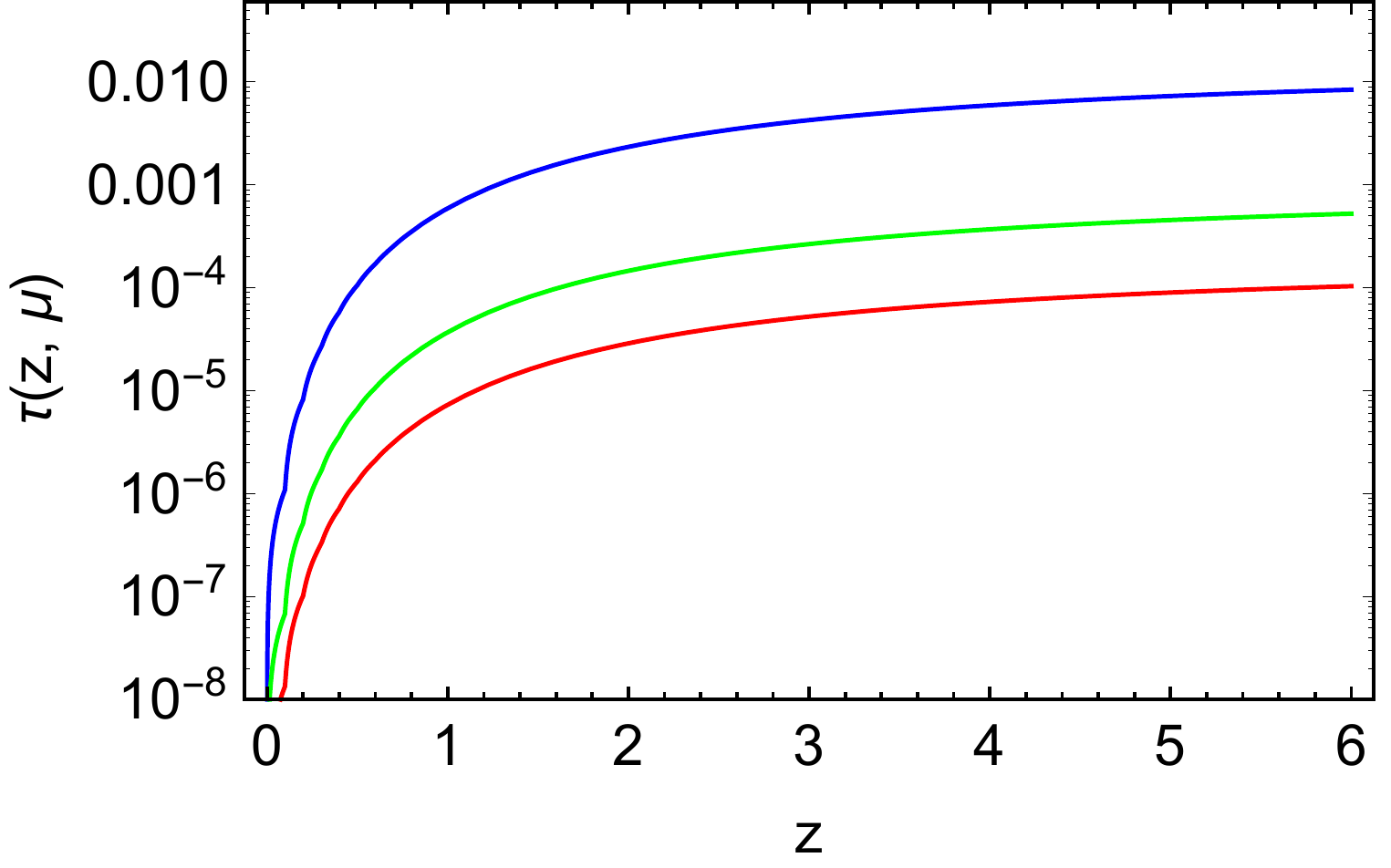}\qquad
\includegraphics[width=0.45\textwidth]{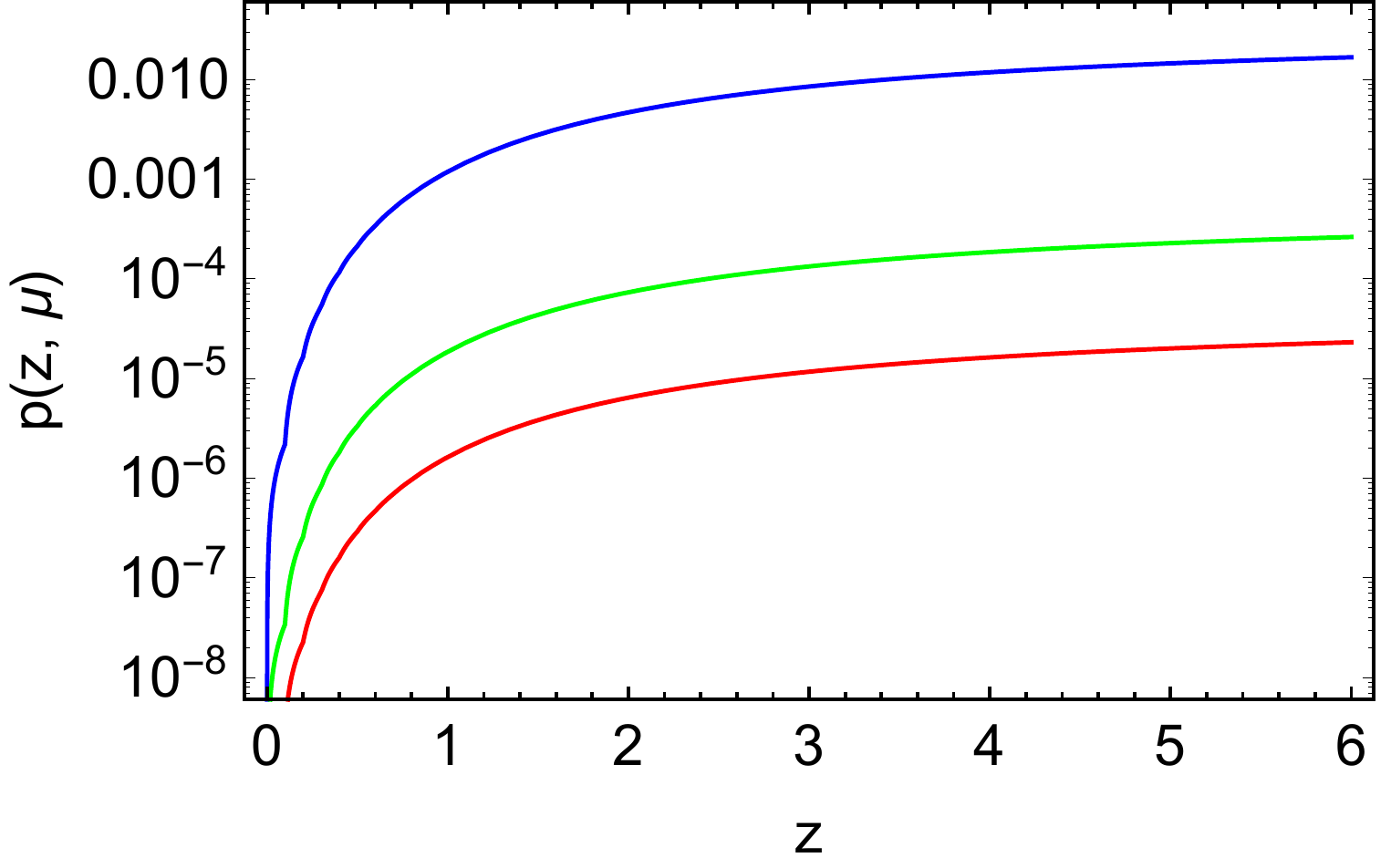} 
 \end{center}
\caption{\label{f:tau-p} Left: Optical depth for $\mu=2, 5, 10$ (from top to bottom) using the lens distribution from \eqref{fit2}.  Right: Lensing probability density $p(\mu, z)$ for $\mu=2, 5, 10$ (from top to bottom) using the lens distribution from \eqref{fit2}. }
\end{figure}

\begin{figure}[ht!]
\begin{center}
\includegraphics[width=0.44\textwidth]{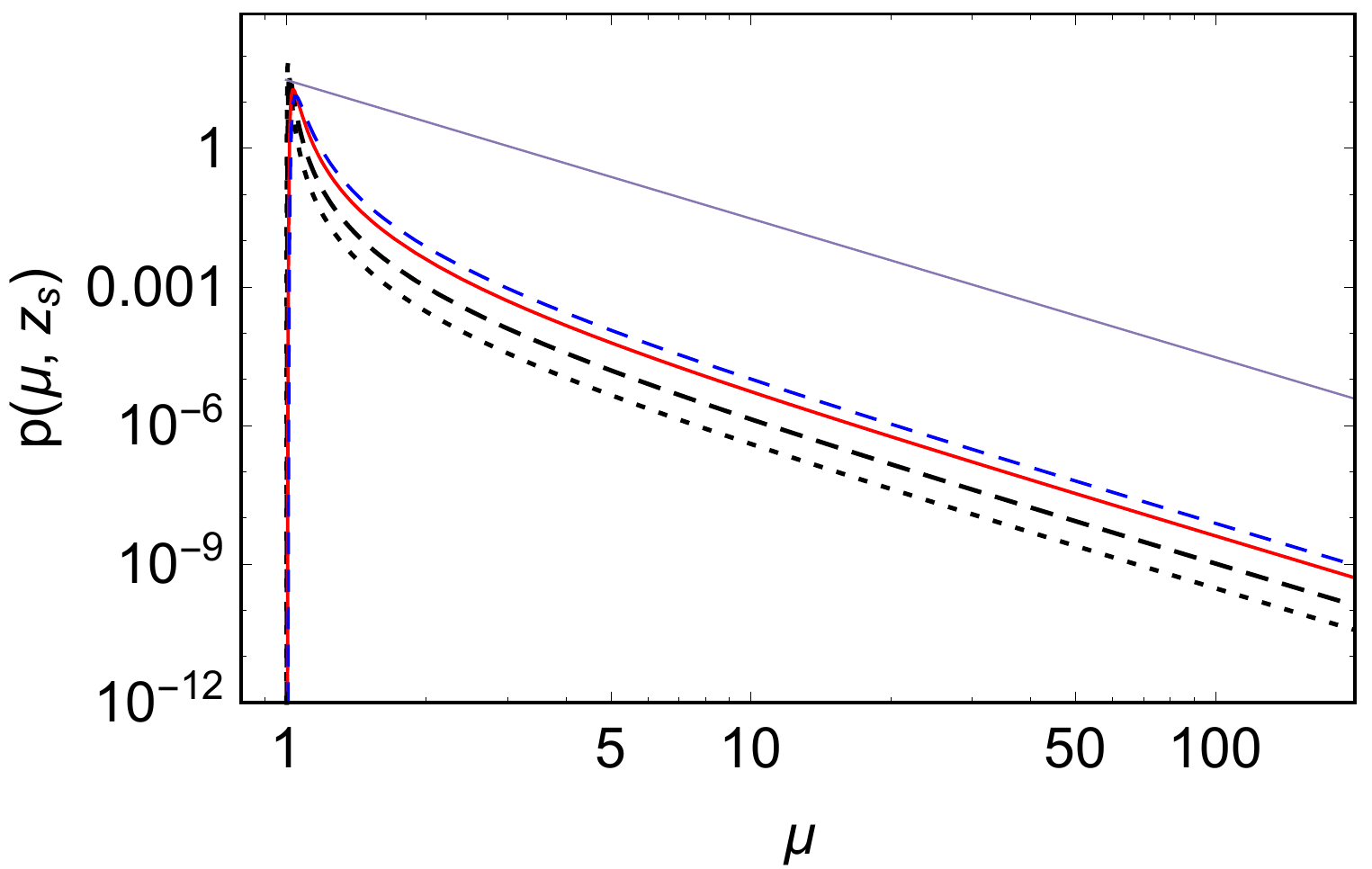}\qquad 
\includegraphics[width=0.44\textwidth]{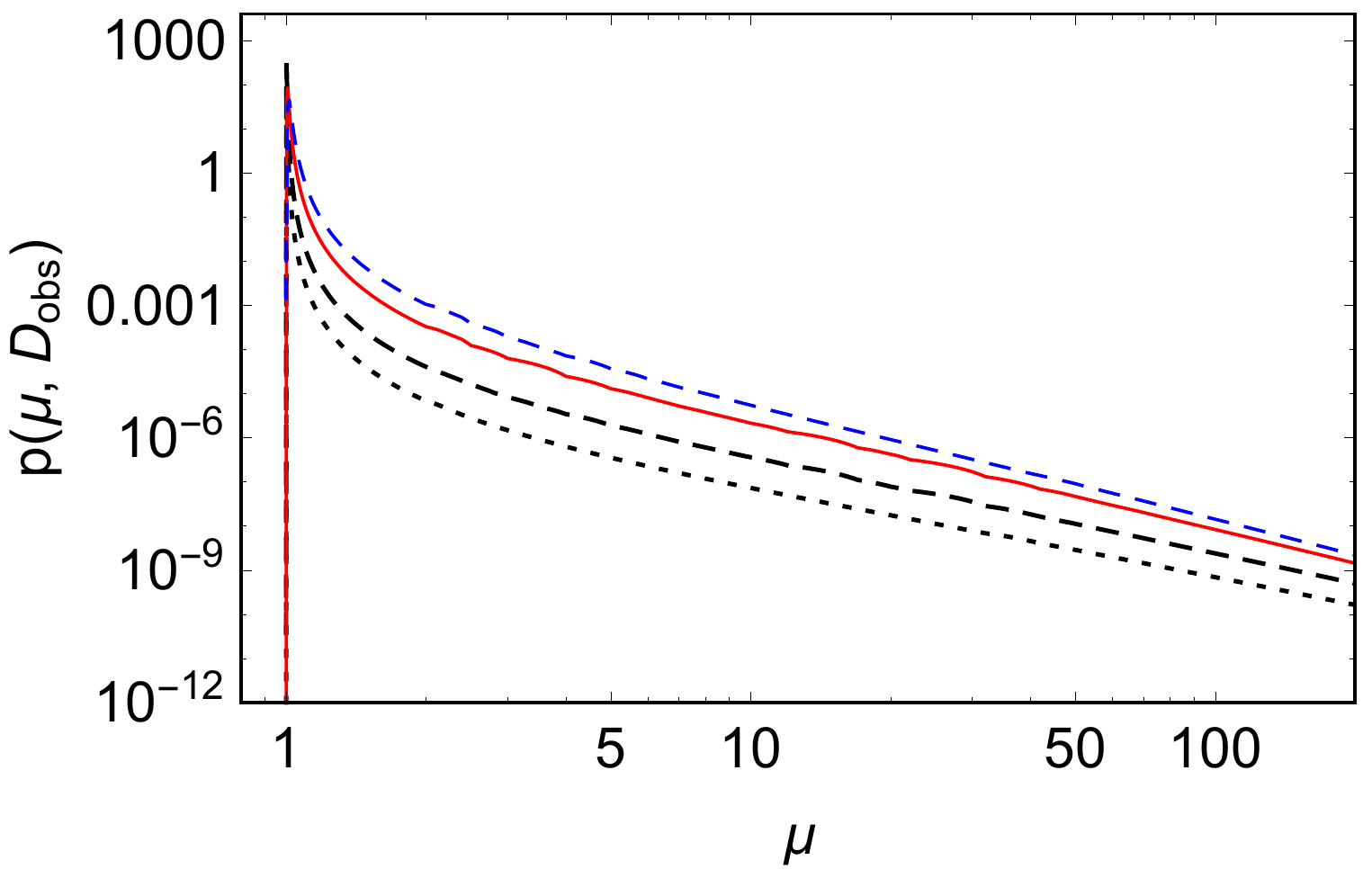}\qquad 
\end{center}
\caption{\label{PP} Left: Lensing probability density $p(\mu, z)$ as a function of magnification for $z=0.6, 1, 2, 3$ (from bottom to top) using the Illustris result \eqref{fit2}.  Right: Lensing probability density $p(\mu, D_{\text{obs}})$ as a function of magnification for $D_{\text{obs}}=500, 1000, 2500, 4500$ Mpc (from bottom to top) using the Illustris result \eqref{fit2}. For sufficiently high redshift  $p(\mu, z)$ scales as $\mu^{-3}$ as mentioned in the text. The thin solid line on the left panel illustrates the scaling.}
\end{figure}

To determine the optical depth given in \eqref{e:depth2}, we need to compute the integral
\be
\int d\sigma_v \sigma_v^4 n(\sigma_v, z)=(1+z)^{3}\int d\sigma_v \sigma_v^4 n_{\text{com}}(\sigma_v, z)\,,
\ee
where we  recall that $\sigma^4_{v}$ is in units of (km/s)$^4\sim 1.2\cdot 10^{-22}$. With this we can compute the cross-section, the optical depth and probability of lensing.  In Fig.~\ref{f:tau-p} we plot the optical depth for $\mu=2, 5, 10$ as a function of redshift on the left panel and the probability density as a function of redshift on the right panel. The probability for lensing magnification, even by just a factor of 2, always remains below 1\%. 
In Fig.~\ref{PP} we plot the probability density as a function of the magnification for different redshifts. On the left hand panel we present curves for fixed source redshifts while in the right hand panel we fix the observer distance $D_{\text{obs}}=D(z_s)/\sqrt{\mu}$. We have verified that the integral of each curve is one, hence the probability density is correctly normalized. 
Note that the probability distribution is always strongly peaked at $\mu\simeq 1$.  We stress that both optical depth and probability density depend only on the lens model and on the model for the distribution of lenses, but they are observatory-independent and they do not depend on the distribution of sources. 

\begin{figure}[ht!]
\begin{center}
\includegraphics[width=0.45\textwidth]{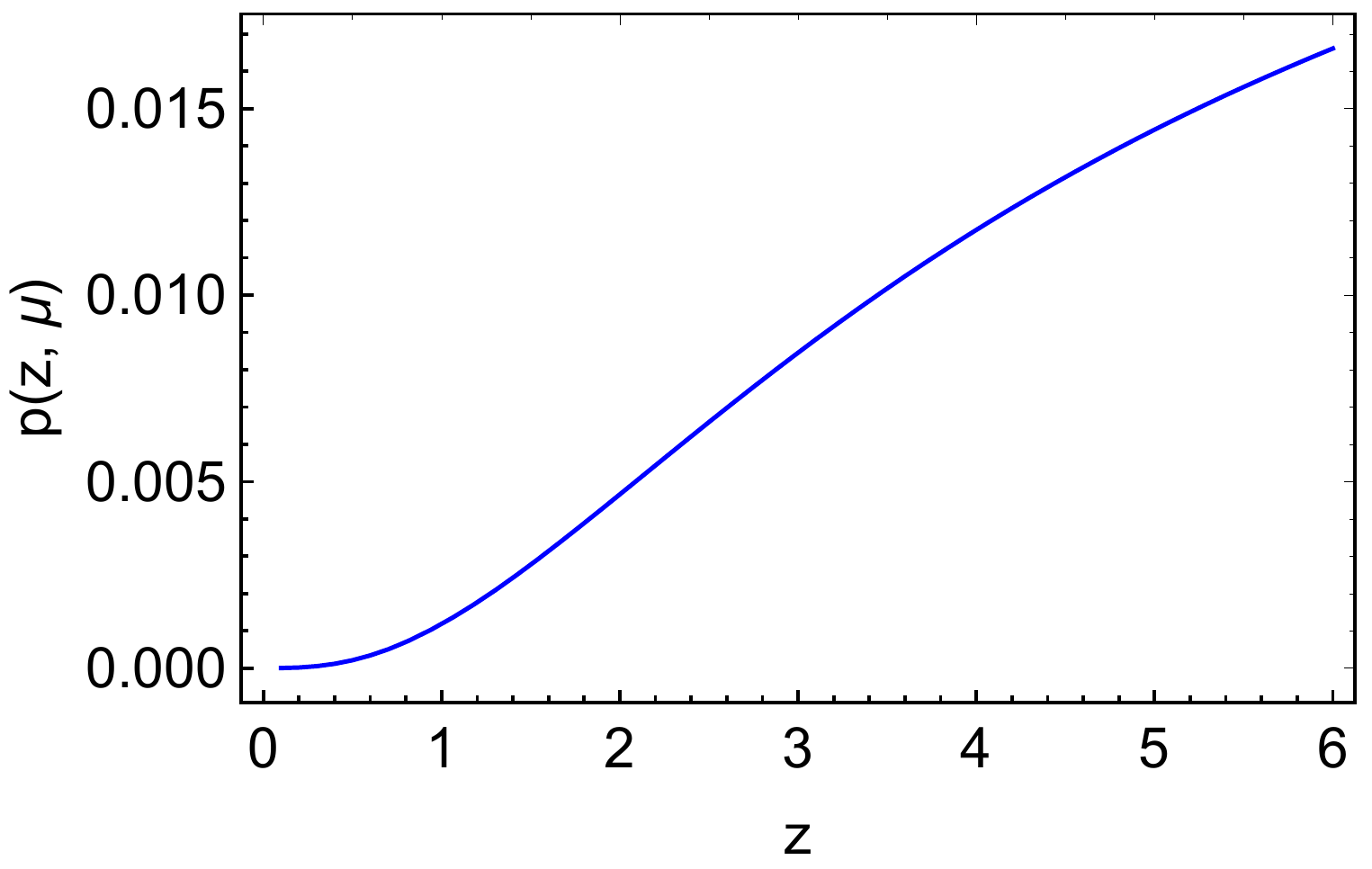} \qquad
\includegraphics[width=0.44\textwidth]{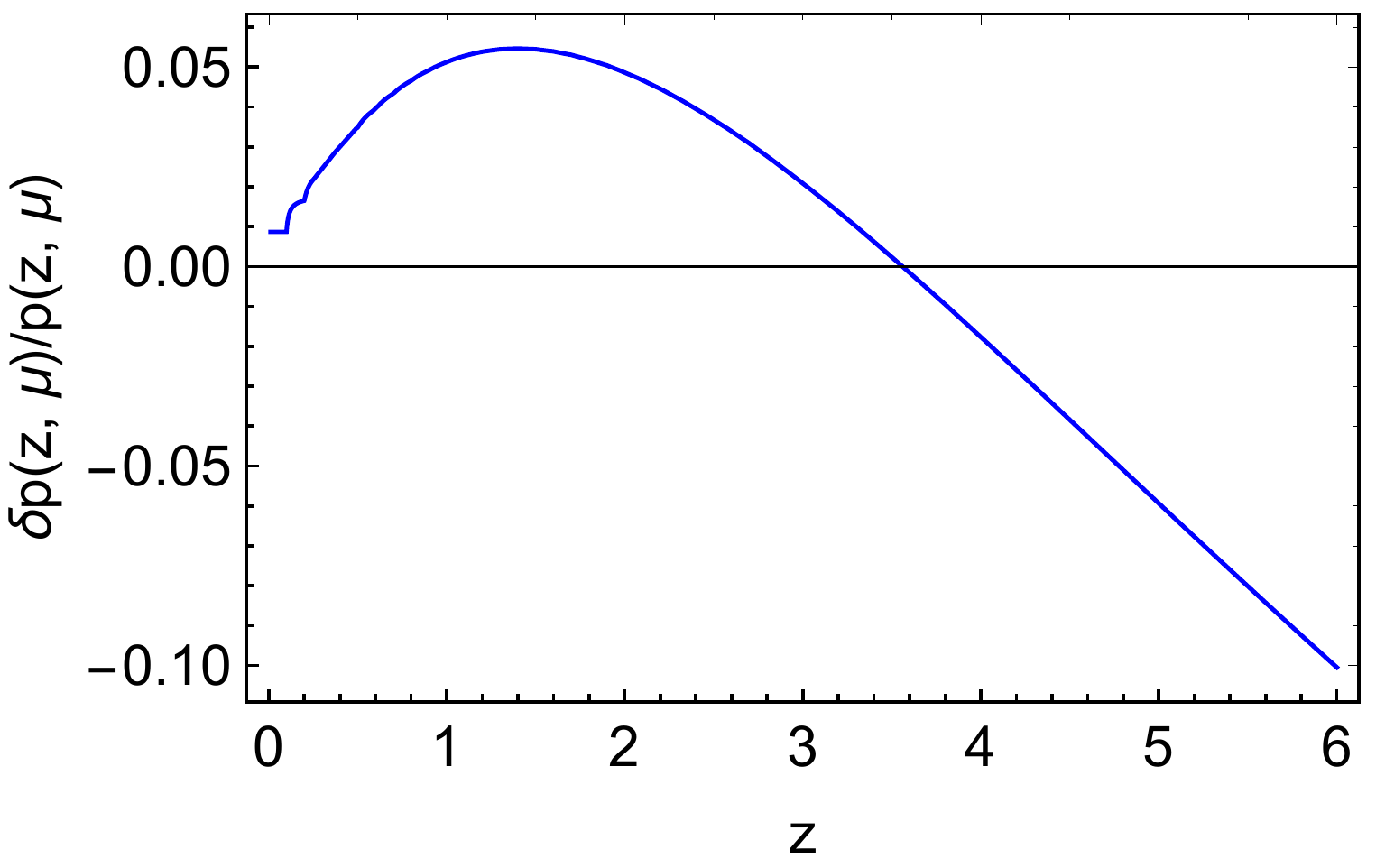} 
 \end{center}
\caption{\label{Pmu} Left : The lensing probability density $p(\mu, z)$ as a function of redshift for $\mu=2$ using the Illustris result \eqref{fit2}. 
Right: The relative difference between the probability with  and without redshift evolution. The result is valid for arbitrary magnification $\mu$. }
\end{figure}

To have some intuition about the role of evolution we finally compare our result with a non-evolving lens distribution. For this we use the distribution function \eqref{fit2} for $z=0$  assuming no redshift evolution. We then have
\be
n^{\text{no evo}}(\sigma_v, z)=(1+z)^{3}n_{\text{com}}(\sigma_v, z=0)\,. 
\ee
In figure \ref{Pmu} we plot the relative difference between the probability density with evolution as given in \eqref{fit2} and without redshift evolution. The result is valid for arbitrary magnification $\mu$.
At low redshift, $0<z<2$, the non evolving distribution actually has somewhat less galaxies with low velocity dispersion which are the most numerous ones and therefore the evolving distribution leads to a higher probability for magnification than the non-evolving one, $\delta P>0$. At higher redshift, $z>4$ the loss of lenses in the evolving distribution becomes more relevant and $\delta P$ changes sign. But the difference between the evolving and non-evolving distribution is never larger than 8\% for redshift $z\leq 6$. This tells us that our results are quite insensitive to the redshift evolution of the lenses.

 \subsection{Compact binary population: numerical modelling}\label{catalogue}

Let us now describe the modeling of the GW sources. As in section~\ref{an:sources} we concentrate on short-lived sources, such as compact binary coalescences which can be observed with LIGO-Virgo \cite{Abbott:2016blz,Abbott:2016nmj,Abbott:2017vtc,TheLIGOScientific:2016pea,Abbott:2017oio,Abbott:2017gyy,LIGOScientific:2018mvr}. 

We focus below on BBH events and assume that their masses are in the range $[5,50]M_{\odot}$ \cite{LIGOScientific:2018mvr,LIGOScientific:2018jsj}. We further assume that their formation rate is proportional to the cosmic star formation rate (SFR), for which we use the following parametric form \cite{Vangioni:2014axa}:
\begin{equation}
 \Psi(t(z))=A\frac{e^{b(z-z_m)}}{a-b+b\cdot e^{a(z-z_m)}}\,,
\end{equation}
with $A = 0.24 M_{\odot}$/yr/Mpc$^3$, $z_m = 2.3$, $a = 2.2$ and $b = 1.4$. Note that we use this parametrization in the redshift range $[0,20]$ even though it was derived using observations up to $z=9$. We define an efficiency parameter $\epsilon$ to describe the fraction of mass in BBH that merge within the age of the Universe that form out of a given stellar population. Moreover, we assume a delay time $t_{\rm delay}$ between the formation of the progenitor star and the coalescence of the BBH, where the latter is distributed as 
\begin{equation}
 P_d(t_{\rm delay})\propto t_{\rm delay}^{-1}\,,
\end{equation}
in the range $[t_{\rm min},t_{\rm max}]$ with $t_{\rm min}=50$ Myr and $t_{\rm max}$ equal to the age of the Universe. We also assume that the mass of the primary BH is distributed as:
\begin{equation}
 P(m_1)\propto m_1^{-\alpha}\,,
\end{equation}
in the range $[5,50]M_{\odot}$ with $\alpha=2.35$, while the mass of the secondary BH is uniformly distributed in the range $[5,m_1]$ so that $p(m_2)=const.$ in this range and $p(m_1,m_2)=p(m_1)p(m_2)$
\cite{LIGOScientific:2018jsj}. Note that this mass distribution is independent of redshift. We assume zero spins for all the BHs. The merger rate is then given by the following integral over the delay time distribution:
\begin{equation}
 \frac{\dd R}{\dd m_1 \dd m_2 \dd z}=\epsilon T_{\rm obs}\int  \Psi(t(z)-t_{\rm delay})p(m_1,m_2)P_d(t_{\rm delay})\frac{\dd V}{\dd z\dd t_{\rm delay}}\,.
 \label{eq:merger_rate}
\end{equation}

   \begin{figure}[ht!]
\begin{center}
\includegraphics[width=0.682\textwidth]{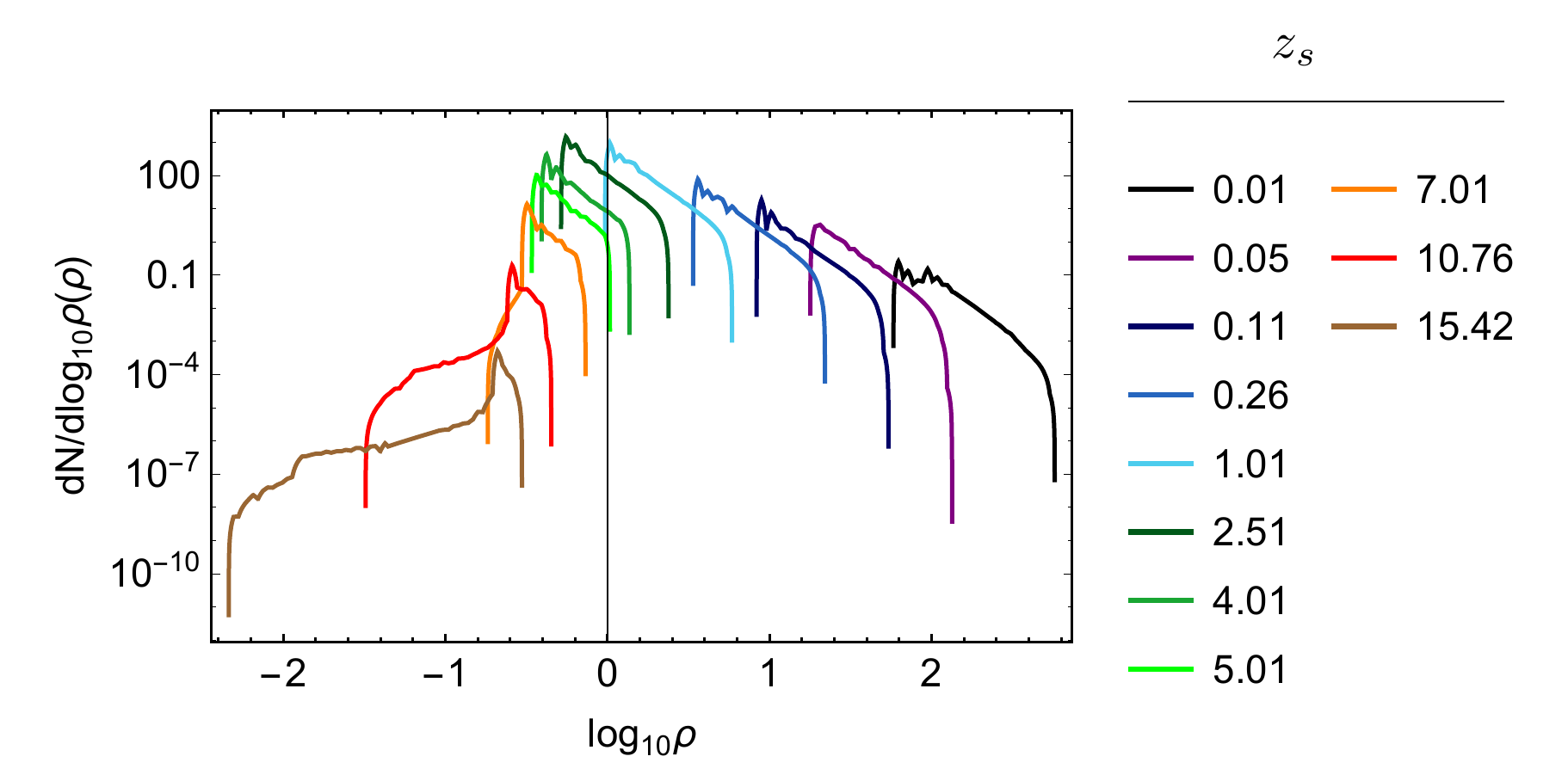} 
\end{center}
\caption{\label{pop} Number of events per unit of $\log_{10}$SNR. Each line corresponds to the number of events in a redshift bin $z\pm0.025$ and the values of redshift are listed in the legend on the right.}
\end{figure}
  \begin{figure}[ht!]
\begin{center}
\includegraphics[width=0.62\textwidth]{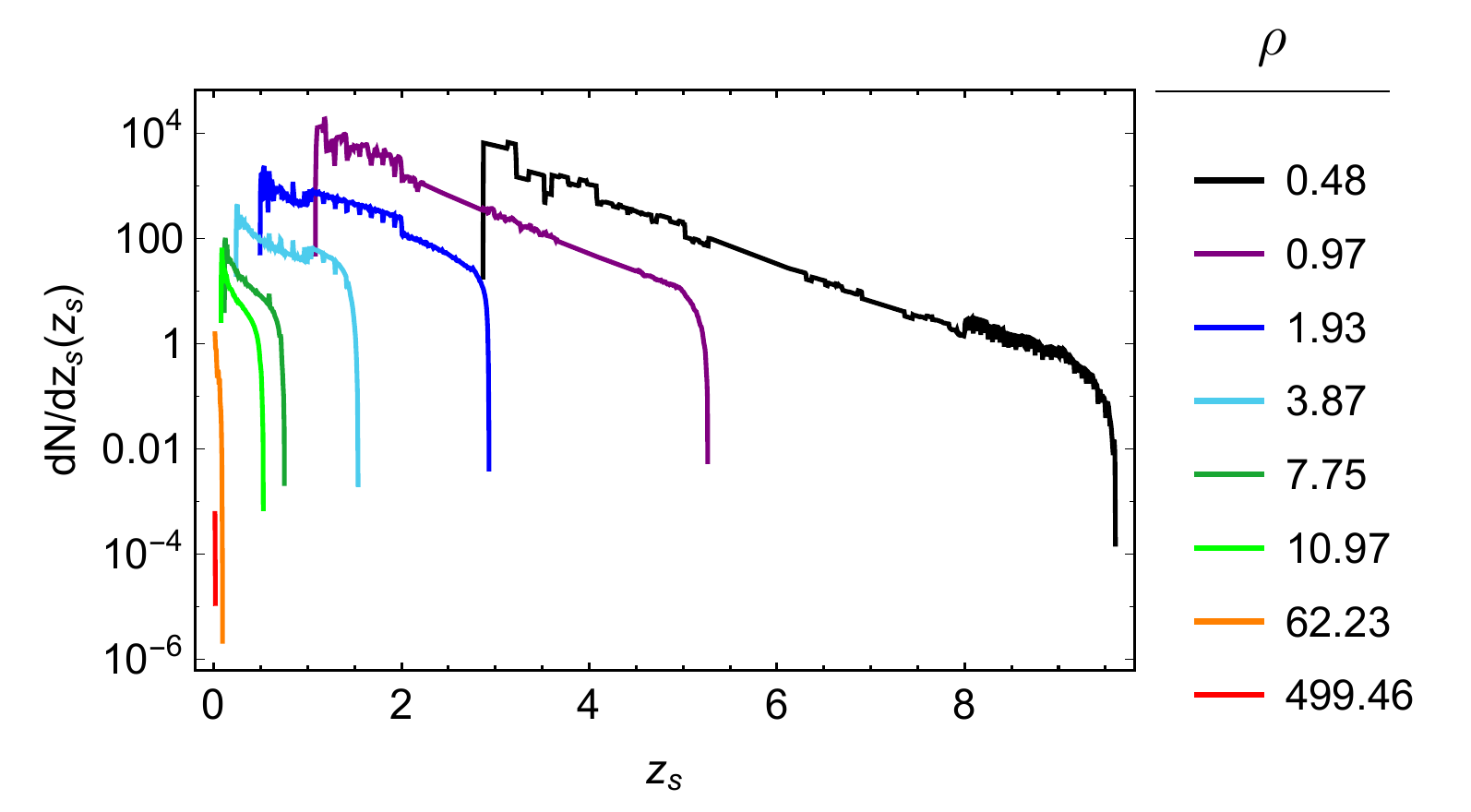} 
\end{center}
\caption{\label{pop2} Number of events per unit of redshift.  Each line corresponds to the number of events in a bin of signal-to-noise  $\log_{10}$SN$ \pm 0.015$ and the values of $\rho$ are listed in the legend on the right of the figure.}
\end{figure}

  \begin{figure}[ht!]
\begin{center}
\includegraphics[width=0.48\textwidth]{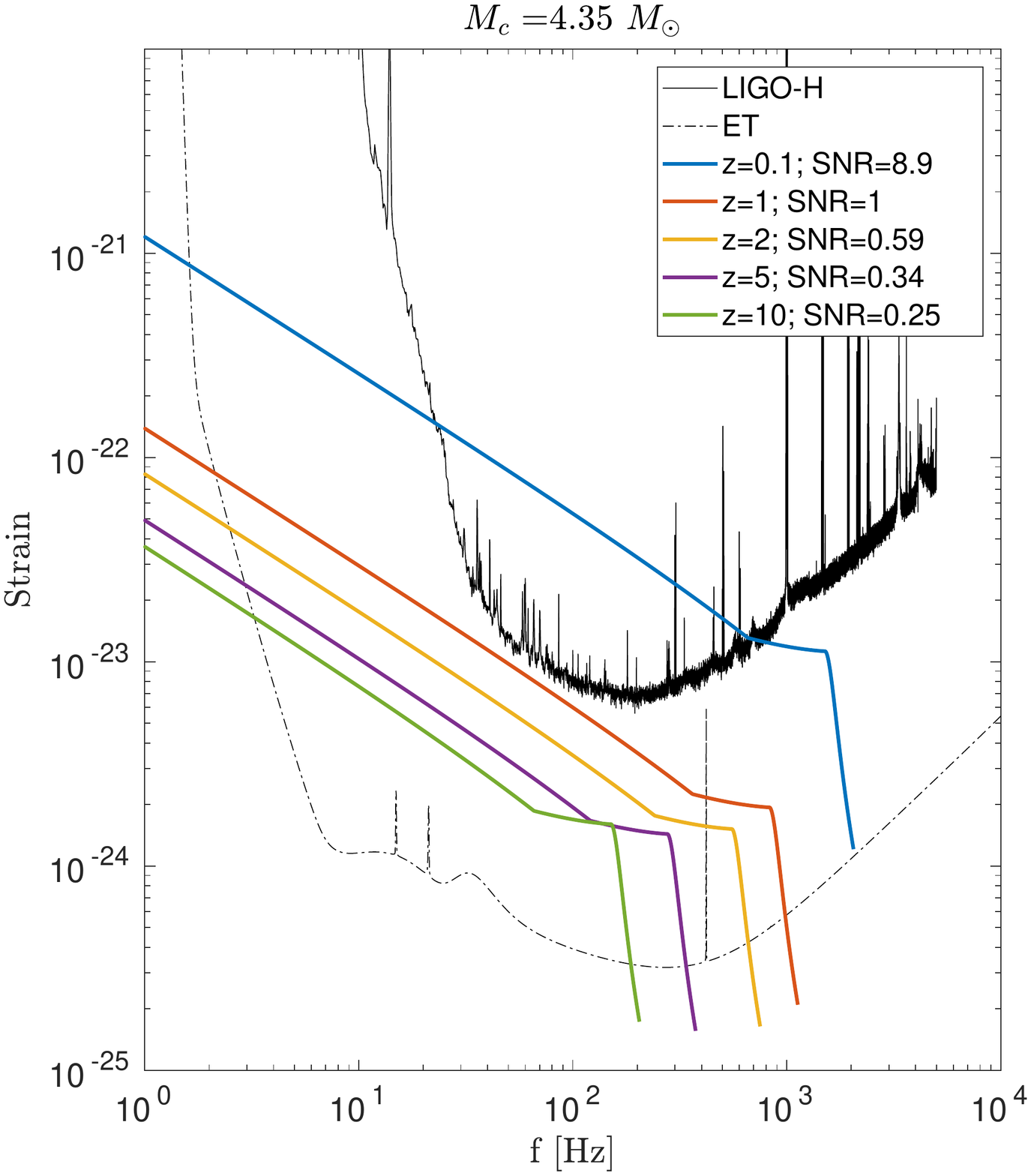} 
\includegraphics[width=0.48\textwidth]{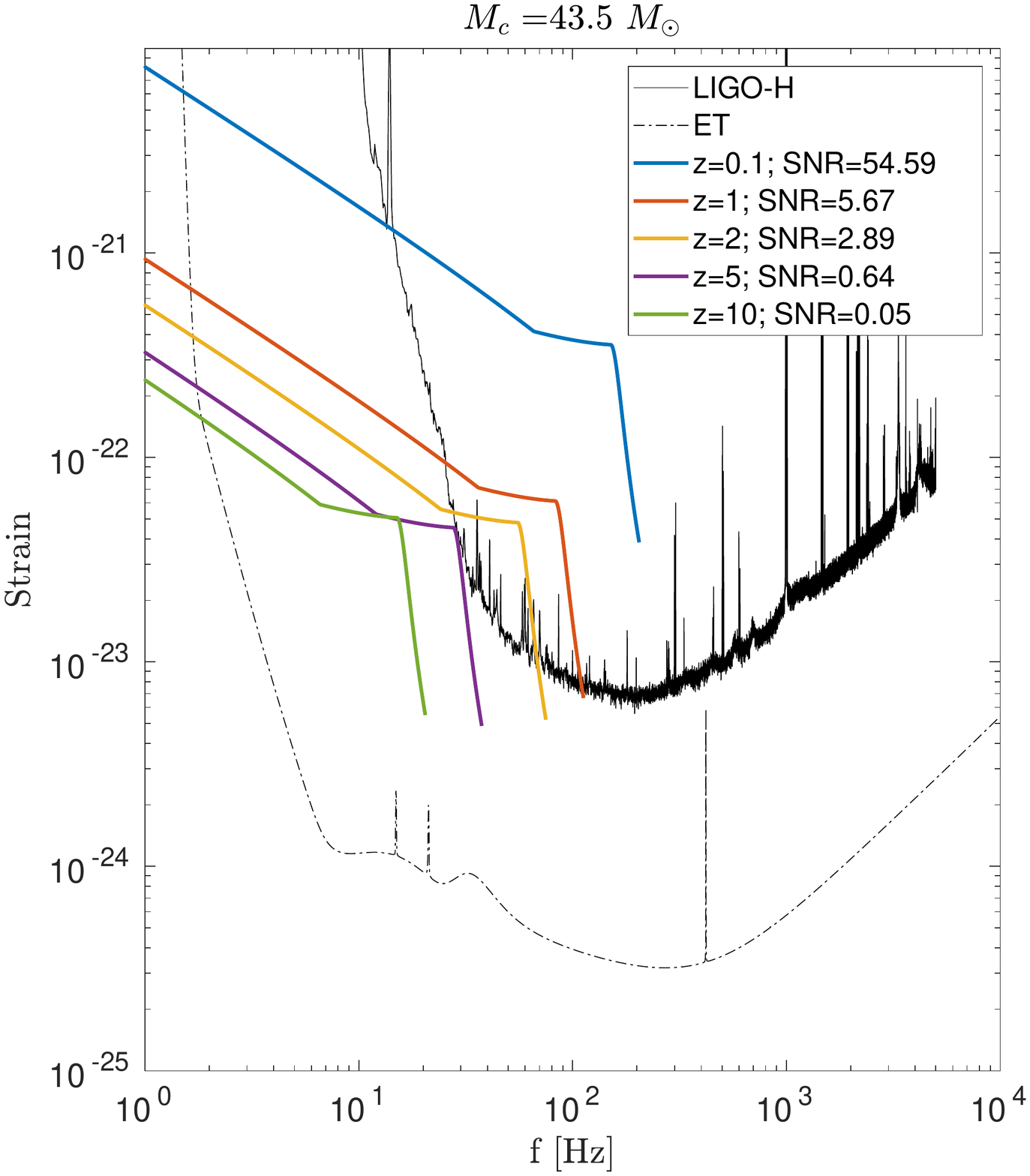} 
\end{center}
\caption{\label{Irina}We fix in each panel a value of the chirp mass and we plot the strain for sources with that mass and in different redshift bins. The solid black curve is the strain sensitivity of the LIGO-Hanford detector \cite{ligo-H1-sen-o1}, the dashed-dotted curve is the strain sensitivity of the Einstein Telescope \cite{ET-sen}.}
\end{figure}

We  calculate the SNR of each merger by using eq.~\eqref{eq:SNR_def}.
 We use the \emph{PhenomB} inspiral-merger-ringdown waveforms to calculate the strain \cite{Ajith:2009bn} and we use the noise power spectral density publicly available from \cite{ligo-L1-sen-o1,ligo-H1-sen-o1}. We consider the event observable if it has $\rho \geq \rho_{\lim}=10$. The normalization parameter $\epsilon$ is obtained by comparing the number of observable events in our model to the number of events observed during the first two LIGO/Virgo observational campaigns O1+O2:
\begin{equation}
 N_{\rm obs} = \int_{\rho(m_1,m_2,z)\geq 10} \frac{\dd R}{\dd m_1 \dd m_2 \dd z} \dd m_1 \dd m_2 \dd z
\end{equation}
where $N_{\rm obs}=10$ and $T_{\rm obs}=170$ days. The merger rate per unit redshift and SNR is then calculated from eq.~(\ref{eq:merger_rate}).

In Fig.\,\ref{pop} we show the number of GW events per unit of SNR in different redshift bins. At low redshift, the slope and the width of the curves are constant, as expected from the analytic analysis of Section~\ref{s:ana}. At high redshift the curves develop a low-SNR tail. The reason for this can be understood from Fig.\,\ref{Irina}  where we plot the characteristic strain of BBH mergers with two different values of the chirp mass (left and right panel) and for different redshifts, together with the LIGO detector noise (black solid) and the ET detector noise (black dashed). At low redshift, the  SNR of high-mass events is always bigger than the one of low-mass events due to their higher amplitude. At high redshift, however, the sharp increase of the noise curve at low frequencies  reduces the SNR of high masses more than the one of low masses which have higher frequencies so that the situation becomes actually reversed and low masses  have a higher SNR. This leads to a spread of the SNR for different events over a large range. Of course for all masses at high redshift the SNR is very small and they can only be observed if they are highly magnified.

 \subsection{Probability of strong lensing for O1+O2 LIGO/Virgo events}\label{s:av}

\begin{figure}[ht!]
\begin{center}
\includegraphics[width=0.43\textwidth]{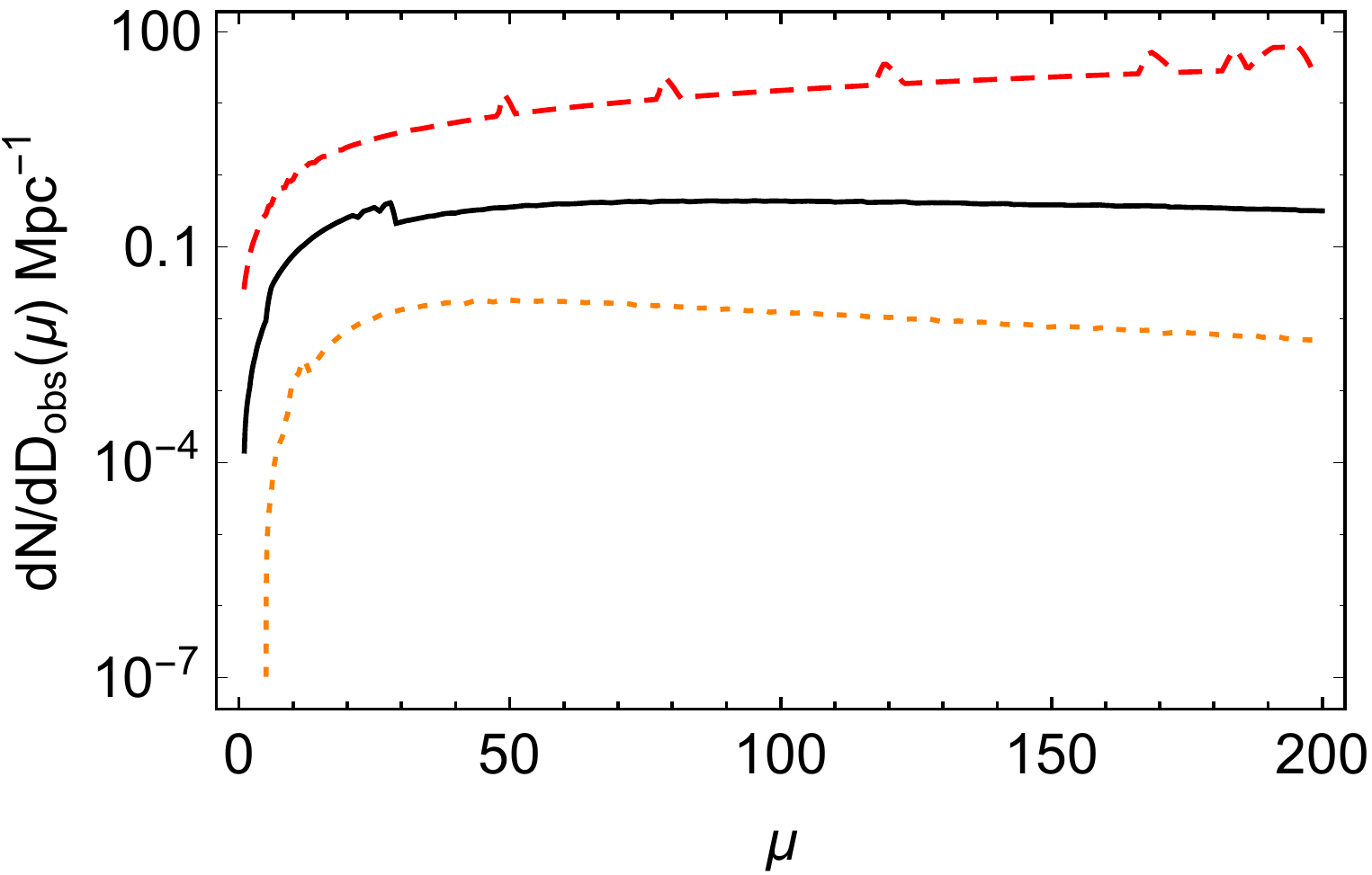}\qquad \includegraphics[width=0.43\textwidth]{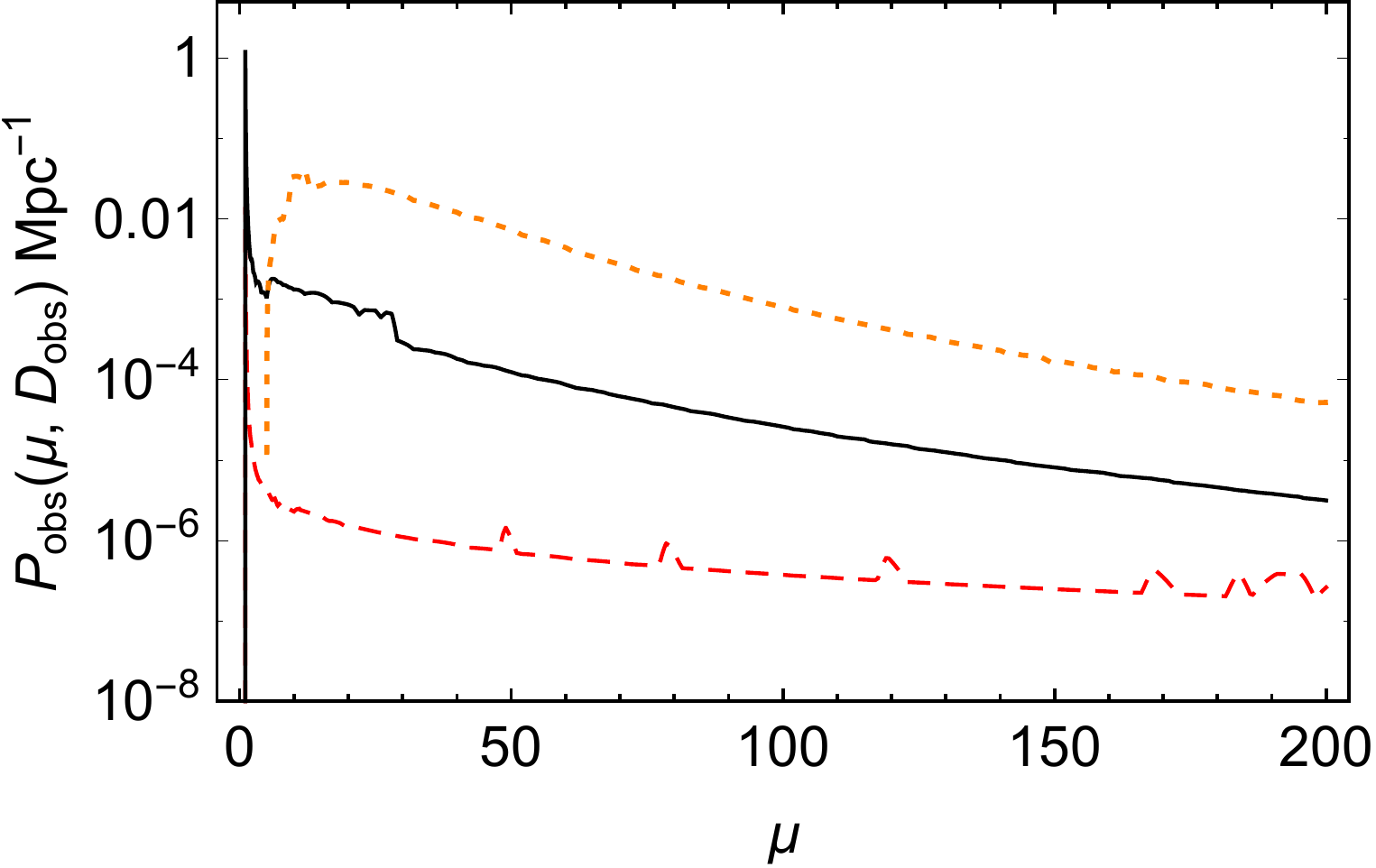}
\end{center}
\caption{\label{Nmu}Left: Number of events per unit of Mpc$^{-1}$ visible if the magnification is $\mu$ as a function of $\mu$, for different observed luminosity distance: $D_{\text{obs}}=(500, 3000, 4500)$ Mpc, red, black and orange curve, respectively.  This quantity is given by $d\mathcal{N}/dD_{\text{obs}}$ in eq.\,(\ref{ForPlot2}).  Right: The same quantity multiplied by $p(\mu, D_{\text{obs}})$ to obtain  the probability distribution of magnification in eq.\,(\ref{PDistr2}) per unit of observed distance.
}
\end{figure}

\begin{figure}[ht!]
\begin{center}
\includegraphics[width=0.43\textwidth]{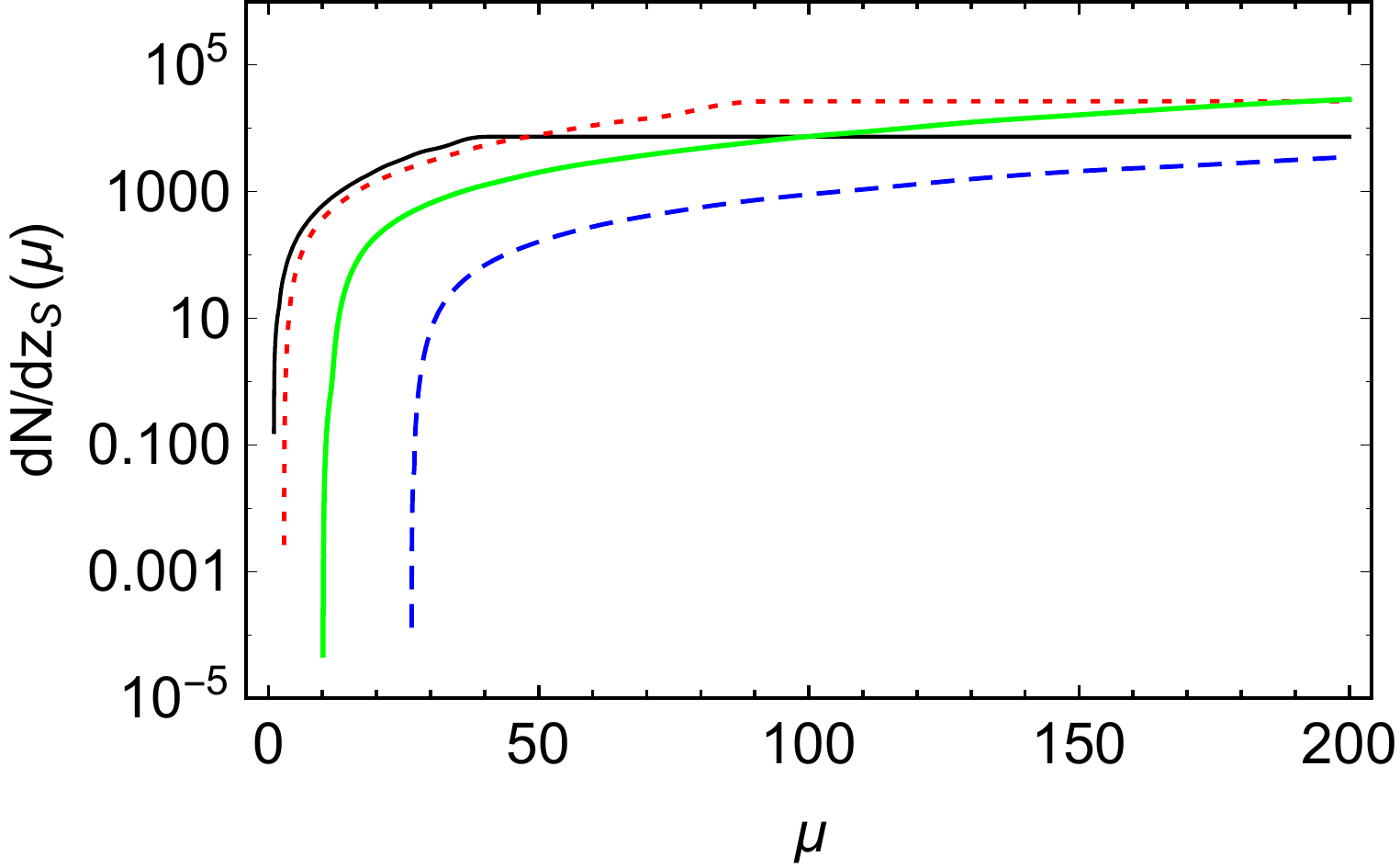}\qquad \includegraphics[width=0.43\textwidth]{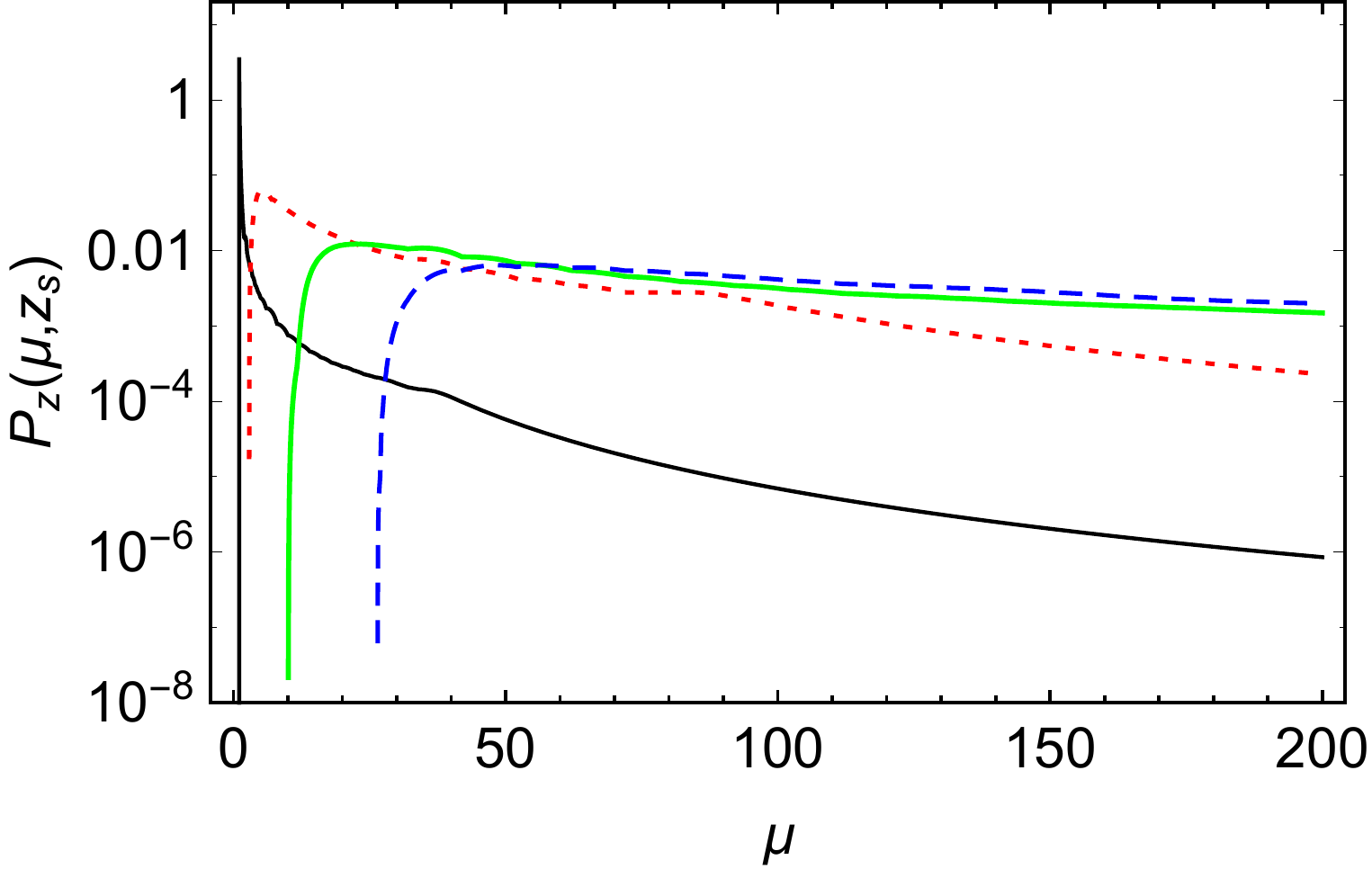}
\end{center}
\caption{\label{NmuZ}Left: Number of events visible if the magnification is $\mu$ per unit of redshift as a function of $\mu$, for different \emph{cosmological} redshift: $z=(0.6, 1, 2, 3)$ black, red, green, blue line respectively. This quantity is given by $d\mathcal{N}/dz_s$ in eq.\,(\ref{ForPlot1}). Right: The same quantity multiplied by  $p(\mu, z_s)$ to obtain  the probability distribution of magnification (per unit of redshift) in eq. (\ref{PDistr}).} \end{figure}

\begin{figure}[ht!]
\begin{center}
\includegraphics[width=0.43\textwidth]{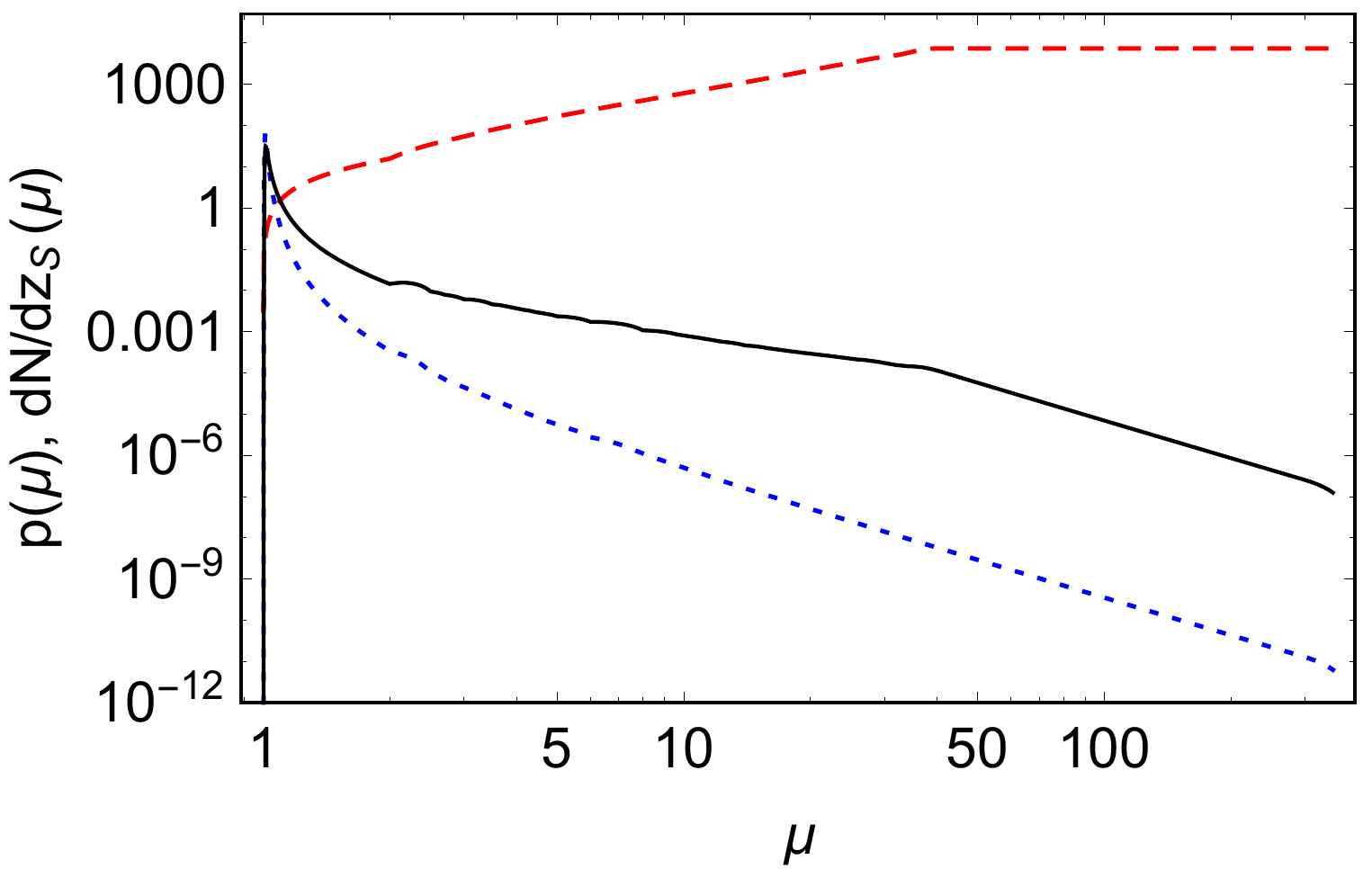}\qquad \includegraphics[width=0.43\textwidth]{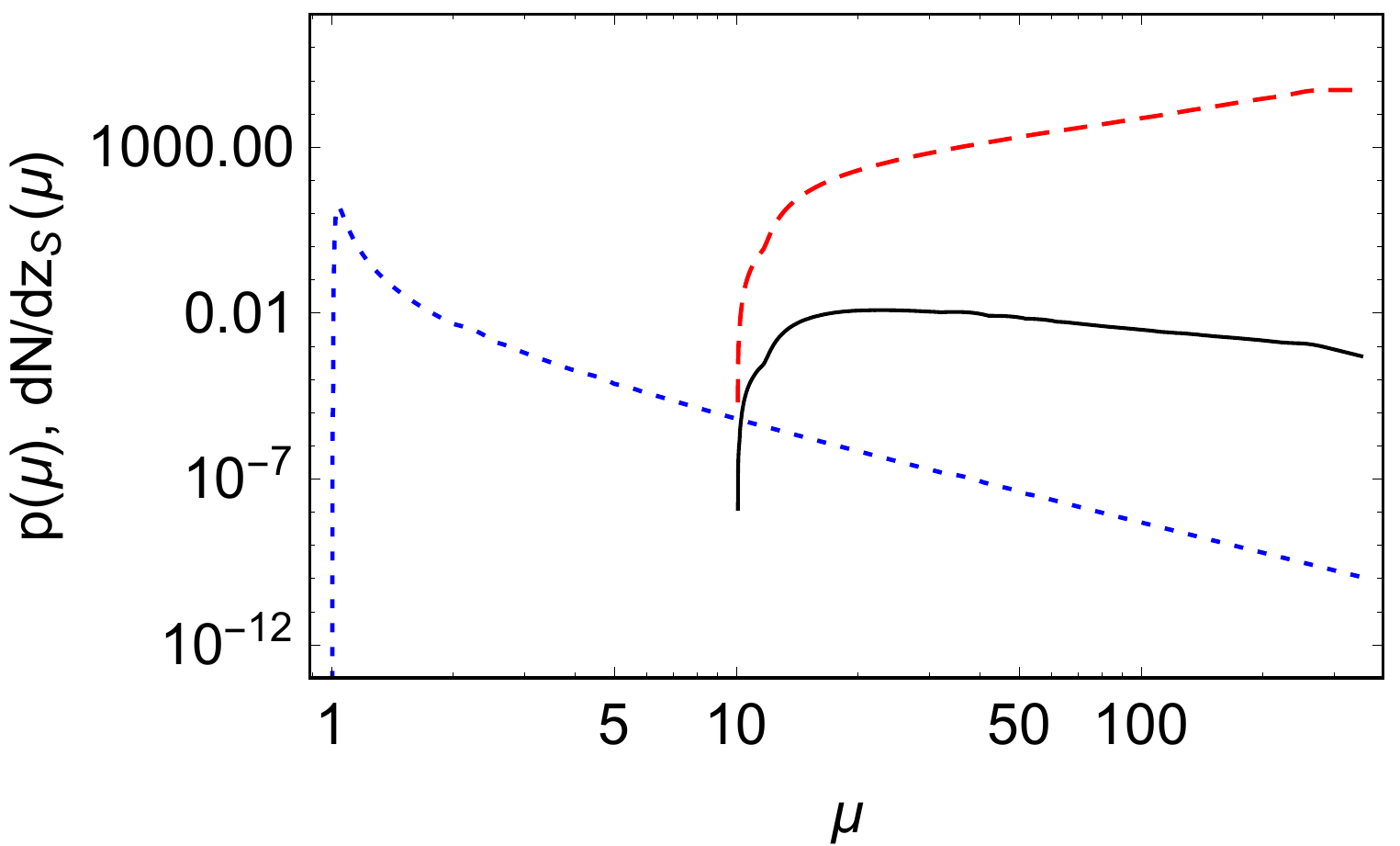}
\end{center}
\caption{\label{clarity} Number of events per unit of redshift visible if the magnification is $\mu$  as a function of $\mu$ (red dashed line), together with $p(\mu, z_s)$ (blue dotted line) and the normalized probability distribution of $\mu$,  $\mathcal{P}_{s}(\mu)$ defined in eq.\,(\ref{PDistr}), proportional to their product (black line). Source redshift: $z_s=0.6$ and $z_s=2$ for the left and right panels, respectively.}
\end{figure}

\begin{figure}[ht!]
\begin{center}
\includegraphics[width=0.43\textwidth]{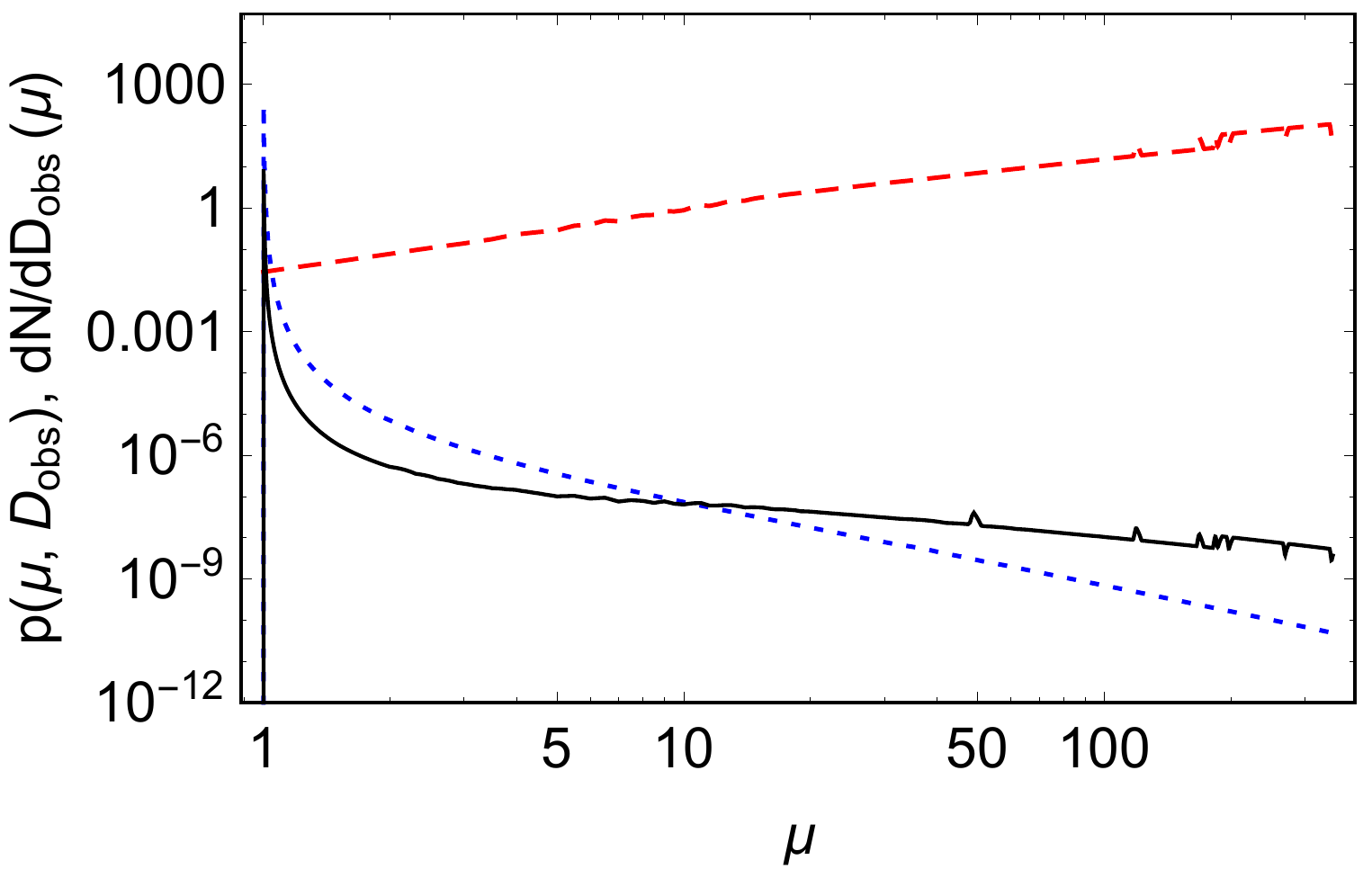}\qquad \includegraphics[width=0.43\textwidth]{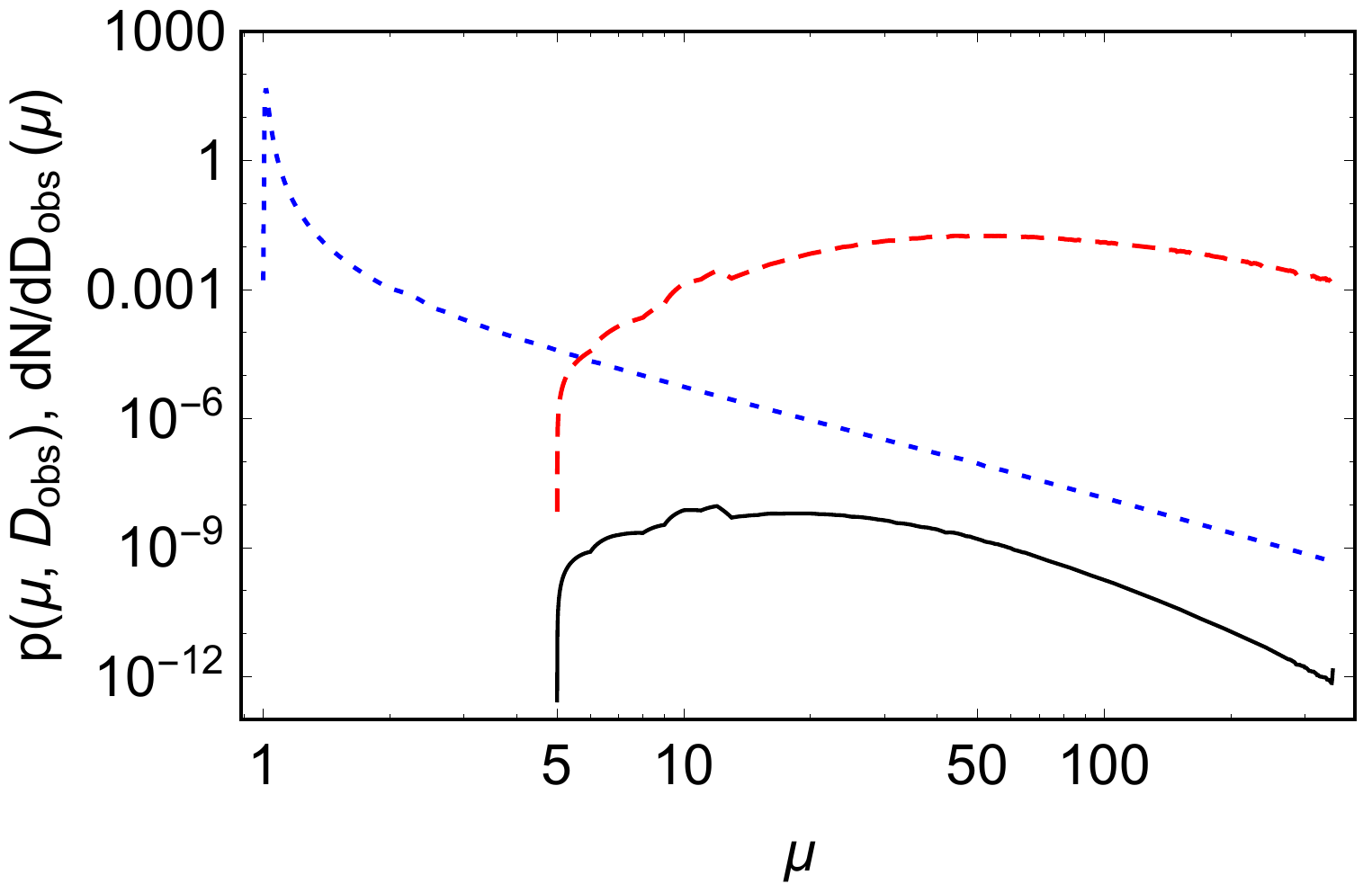}
\end{center}
\caption{\label{clarity2} Number of events per unit of observed distance visible if the magnification is $\mu$ as a function of $\mu$ (red dashed line), together with $p(\mu, D_{\text{obs}})$ (blue dotted line) and the normalized probability distribution of $\mu$, $\mathcal{P}_{\text{obs}}(\mu)$ defined in eq.\,(\ref{PDistr2}),  proportional to their product (black line). Observed distance: $D_{\text{obs}}=500$ Mpc and $D_{\text{obs}}=4500$ Mpc in the left and right panels, respectively. }
\end{figure}

In Fig.\,\ref{Nmu} we present  the number density of magnified objects that we observe from a distance $D_{\text{obs}}$ as a function of magnification. We plot this quantity  for three different values of $D_{\text{obs}}$. In the right hand panel of Fig.\,\ref{Nmu} we multiply each line of the left hand panel by  the probability density of lensing as a function of magnification to obtain the (normalized) probability distribution of magnification for objects observed from different distances, $D_{\text{obs}}$, i.e. eq.\,(\ref{PDistr2}), as a function of magnification. Note that while the number of objects which can be seen by the given survey (with fixed magnification) is somewhat lower for $D_{\text{obs}}=3000$Mpc than for $D_{\text{obs}}=500$Mpc, the probability of seeing such objects is higher for $D_{\text{obs}}=3000$Mpc than for  $D_{\text{obs}}=500$Mpc. This is due to the fact that  $p(\mu,D_{\text{obs}})$ is monotonically increasing with  $D_{\text{obs}}$, see Fig.~\ref{PP}. For the product, $p(\mu,D_{\text{obs}})d{\cal N}/dD_{\text{obs}}$ the increase of $p$ wins over the 
decrease of $\cal N$ for $D_{\text{obs}}=3000$Mpc. 

In Fig.\,\ref{NmuZ} we plot the same quantities for different values of the cosmological redshift of the source. At low redshift, $d{\cal N}/dz$ tends to a constant at high magnification since we see all events (black curve, left panel). At high redshift $p(z,\mu)d{\cal N}/dz$ tends nearly to a constant at high redshift which indicates that  $d{\cal N}/dz$ grows roughly like $\mu^3$ at high redshift (blue dashed curve, right panel). 

\begin{figure}[ht!]
\begin{center}
\includegraphics[width=0.47\textwidth]{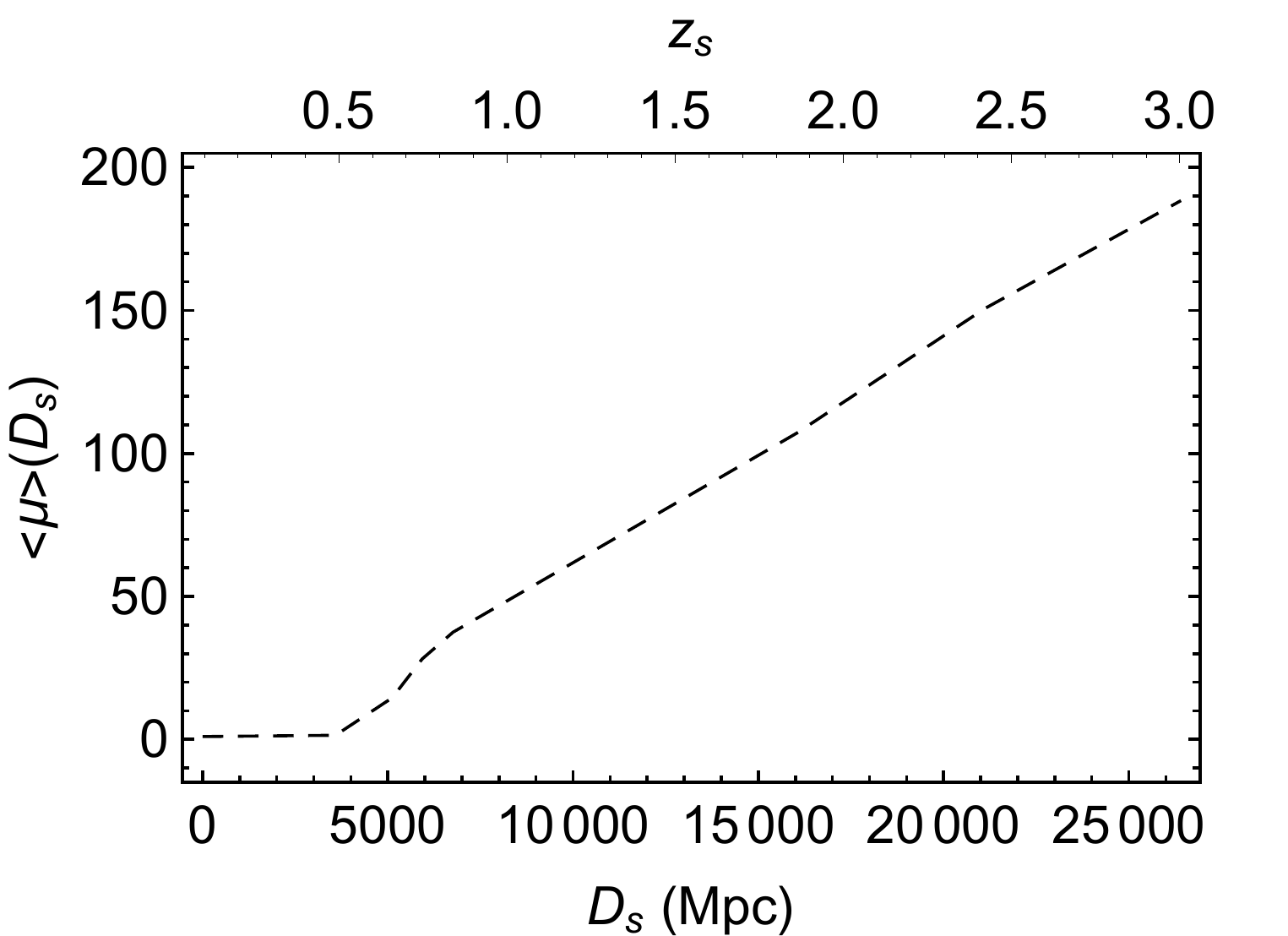}\qquad \includegraphics[width=0.47\textwidth]{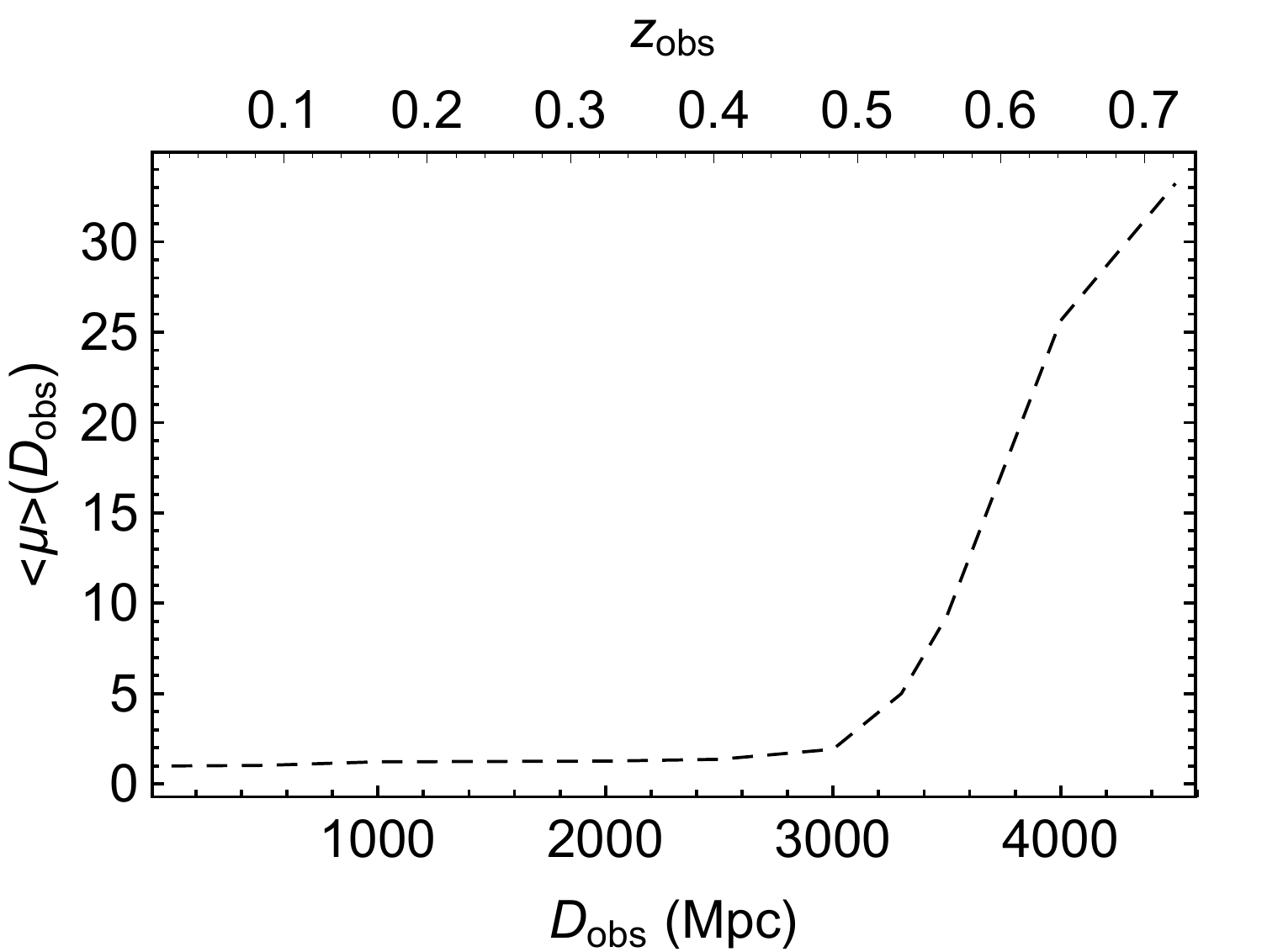}
\end{center}
\caption{\label{mu2} Left: Amplification factor for events observable by LIGO-Virgo, as a function of \emph{cosmological} redshift (and luminosity distance) of a source. Right: Amplification factor for events observable by LIGO-Virgo, as a function of \emph{observed} redshift (luminosity distance) of a source. As explained in the text, no events are observed after 5500 Mpc, independent of the value of magnification. }
\end{figure}

\begin{figure}[ht!]
\begin{center}
\includegraphics[width=0.47\textwidth]{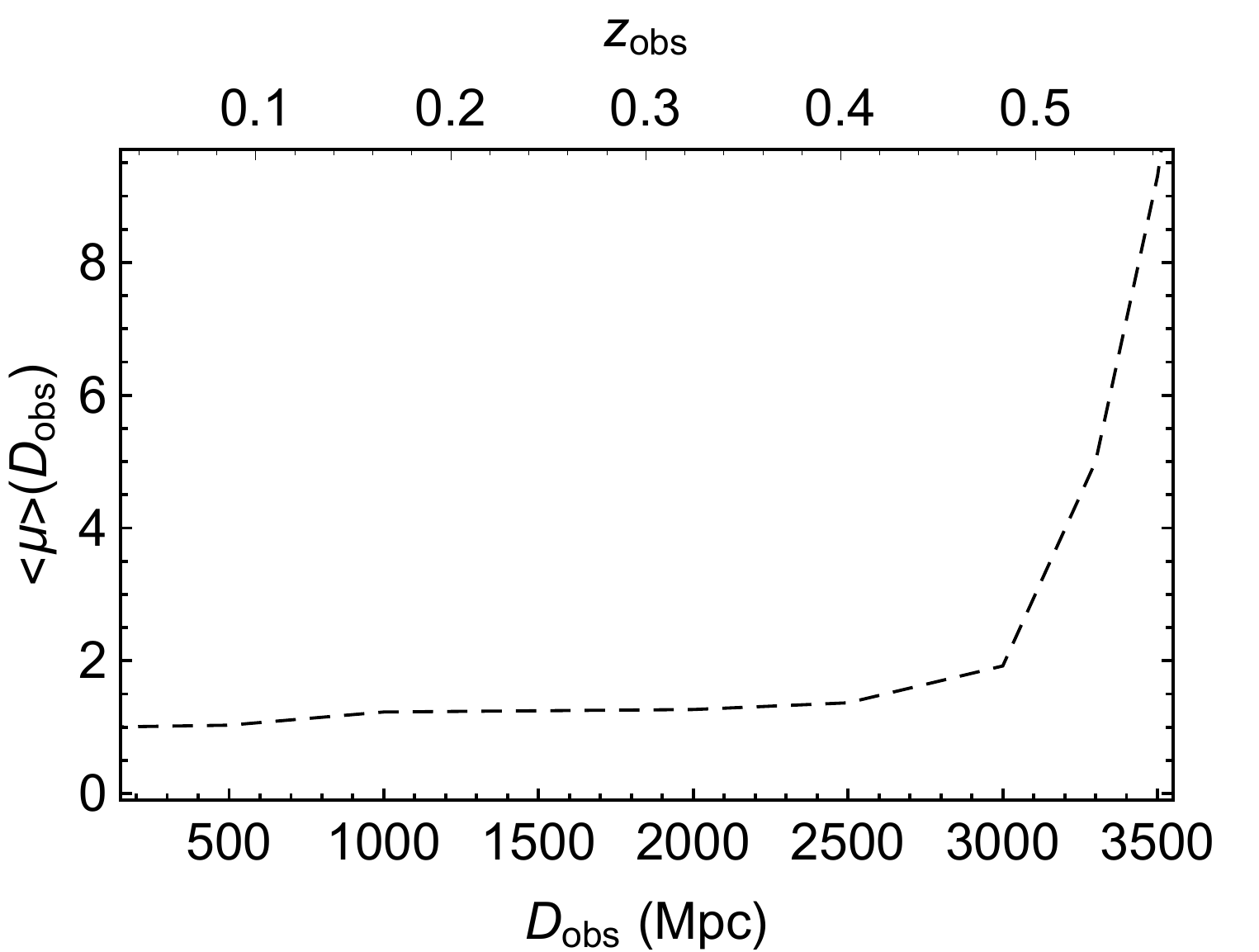}\,
\includegraphics[width=0.47\textwidth]{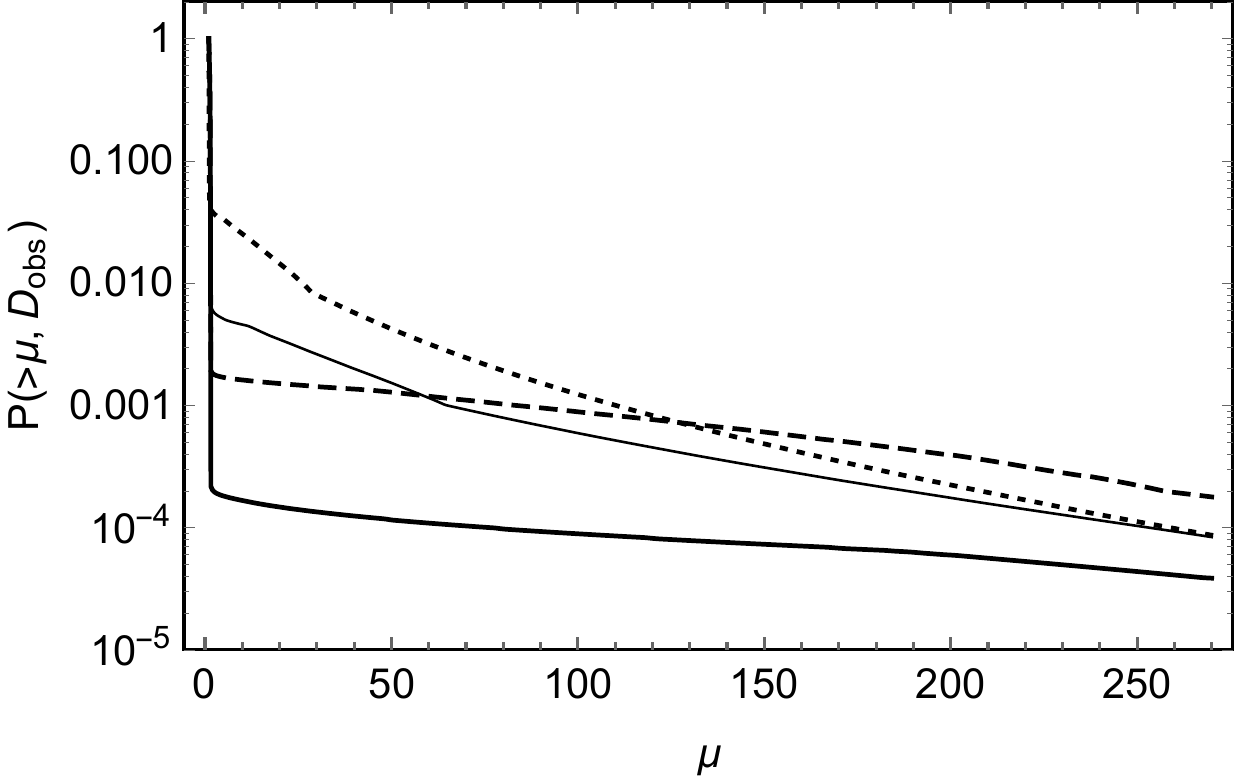}
\end{center}
\caption{\label{PmuInt} Left: Amplification factor for events observable by LIGO-Virgo, as a function of \emph{observed} redshift (or luminosity distance) of a source. This is a zoom to low redshift of the right panel of Fig.\,\ref{mu2}. Right: Probability that a given GW event observed by LIGO-Virgo has magnification bigger than a given value $\mu$, as a function of $\mu$,  for different observed distances $D_{\text{obs}}=500, 1000, 2000, 3000$, from bottom to top respectively.}
\end{figure}

In Fig. \ref{clarity} we show (black line) the probability distribution of magnification of the observed sources (\ref{PDistr}) as a function of magnification, fixing (in each panel) a value of the cosmological redshift of the source. We see that the peak of the distribution moves towards high values of the magnification as we increase the cosmological redshift (from the left to the right panel). In each panel we also plot the number density of sources that we observe from the given observed distance if the magnification is $\mu$ (red line) and the probability density of lensing as a function of magnification (blue line). The (not normalized) distribution of magnification (black like) is the product of these two quantities. In Fig.\,\ref{clarity2} we present similar results but now fixing the observed distance $D_{\text{obs}}$ (in each panel). 
In Fig.\,\ref{mu2} we present the results for the average  amplification factor as a function of cosmological redshift and luminosity distance of the sources (left panel) and as a function of observed redshift and luminosity distance of the sources (right panel). No sources are observable beyond 5500 Mpc.\footnote{This can be understood computing the probability of seeing a source from a given observed distance, i.e. integrating $p(\mu, D_{\text{obs}})$ over $\mu$. The probability of seeing an object from distance bigger than $5000$ Mpc is smaller than $10^{-7}$.} 
The minimum amplification needed to detect sources from a redshift $z_s$ and with a given $\rho_{\text{max}}(z_s)> \rho_{\text{lim}}$ is given by
\be
\mu_{\text{min}}(z_s)=\left(\frac{\rho_{\text{lim}}}{\rho_{\text{max}}}\right)^2\,,
\ee
hence we can see sources from a maximum observed distance given by
\be
D_{\text{obs}}^{\text{max}}(z_s)=\frac{D(z_s)}{\sqrt{\mu_{\text{min}}}}=\frac{D(z_s)\rho_{\text{max}}(z_s)}{\rho_{\text{lim}}}\,. 
\ee
For low redshifts, $z_s\lesssim 2$  we can work out the scaling of this quantity with redshift from our simple model just using eq. (\ref{max}). We find
\be
D_{\text{obs}}^{\text{max}}(z_s)= \frac{\mathcal{A}}{\rho_{\text{lim}}}\left(\frac{\mathcal{M}_{\text{max}}}{M_{\odot}}\right)^{5/6}(1+z_s)^{5/6}f_{7/3}(z_s)^{1/2}\text{Mpc}\,.
\ee
For higher redshift our analytical model for the source distribution becomes too inaccurate and this scaling is no longer valid.  The left panel of Fig.\,\ref{PmuInt} is a zoom of the low redshift part of Fig.\,\ref{mu2}. The probability,  $\mathcal{P}_{\text{obs}}(>\mu)$ that an object that LIGO-Virgo observes from a distance $D_{\text{obs}}$ has been magnified more than $\mu$  is presented in Fig.\,\ref{PmuInt} for different values of $D_{\text{obs}}$, as a function of $\mu$ (right panel).

\section{Discussion and conclusion}\label{s:con}

Due to cosmological expansion, the frequencies of gravitational waves are redshifted. Since the frequencies from BBH mergers are degenerate with the masses, the quantity that we measure directly from the signal is the redshifted chirp mass $\mathcal{M}_z = (1 + z)\mathcal{M}$, where $\mathcal{M}$ is the intrinsic (true) chirp mass of the binary  at a redshift $z$. The estimated luminosity distance can be converted into a redshift estimate (by choosing a cosmological model), which can, in turn be used to estimate the intrinsic chirp mass $\mathcal{M}$ of the binary. The unknown lensing magnification will bias our estimation of the intrinsic mass and the distance (or equivalently, the redshift) of the binary. Therefore, lensed binaries will appear as a population at a lower redshift and with higher masses. 

In this article we have proposed a semi-analytical approach to study weak and strong lensing of gravitational waves by galaxies. We presented a fully analytical model to study the dependence of the result on the choice of the distribution of lenses and  on the properties of the source population. The framework we propose can be applied to any detector or detector network once the strain noise of the observatory is known. We also developed a more complete numerical study to include a realistic evolution of the lenses with time. We presented results for the probability of strong lensing as a function of redshift and magnification and for the average amplification as a function of cosmological redshift and observed distance. 

Results for the probability that a source with cosmological redshift $z_s$ is magnified by $\mu$ are presented in Fig.\,\ref{PP}, as a function of both cosmological and  observed redshift (or rather distance) of a source. We stress that this result does not depend on the properties of the distribution of sources, nor on the specifics of a given observatory. The probability is small, but it decays slowly as a function of magnification. More specifically, for large values of the magnification, the probability as a function of cosmological redshift scales as $p(\mu, z_s)\propto \mu^{-3}$. This scaling is independent of  the lens distribution model. The probability density as a function of the observed distance, $p(\mu, D_{\text{obs}})\equiv p(\mu, z_s(\sqrt{\mu}D_{\text{obs}}))$ has an even  slower decay with magnification, see right panel of Fig.\,\ref{PP}. Indeed, for a fixed $D_{\text{obs}}$ and $\mu>1$,  $p(\mu, D_{\text{obs}})\equiv p(\mu, z_s(\sqrt{\mu}D_{\text{obs}}))$ takes contributions  from a higher redshift than $z_s(D_{\text{obs}})$ and is therefore larger.  The exact scaling of $p(\mu, D_{\text{obs}})$ with $\mu$ depends on the details of the lens distribution. For a realistic distribution with a comoving number density of lenses which fits numerical simulations  (Illustris), we find that $p(\mu, D_{\text{obs}})\propto \mu^{-\alpha}$. The value of $\alpha$ increases from 2 to 3 as  the observed luminosity distance varies from $\sim 300$ Mpc to $5000$ Mpc where it saturates to the asymptotic value $\alpha=3$. From Fig.\,\ref{PP} we read the probability that a source with  observed luminosity distance of $D_{\text{obs}}=2500$ Mpc is magnified by $\mu=2$,  $p(2,2500\text{Mpc})\simeq 10^{-3}$. The probability that a source observed from this same distance is magnified by $\mu=10$ is $10^{-5}$. Notice that this result is detector-independent.

Selection effects related to the sensitivity curve (hence to the horizon) of a given detector network enter into the game when computing the probability that a source \emph{that we can observe} with a given observatory has a given magnification. This probability distribution $\mathcal{P}_{\text{obs}}(\mu)$, is defined in eqs.\,(\ref{PDistr}) and (\ref{PDistr2}) as  function of the cosmological redshift of a source and as  a function of the observed luminosity  distance, respectively. We focussed on the LIGO-Virgo network in our analysis. Plots of the distribution of magnification for different cosmological redshifts and observed luminosity distance are presented in Figs.~\ref{clarity} and \ref{clarity2} (black lines). The distribution  is highly peaked at $\mu=1$ for low observed distances but it has more and more power at higher magnification as we move towards higher redshift or observed distance. 
The integral of this distribution multiplied by  $\mu$ gives the average magnification expected for a given observed distance and the given observatory.

The probability distribution of magnification $\mathcal{P}_{\text{obs}}(\mu)$ for a given observed distance scales as $\mathcal{P}_{\text{obs}}(\mu)\propto \mu^{-\alpha+1/2}$ and its integral from $\mu$ to infinity behaves as  $\mathcal{P}_{\text{obs}}(>\mu)\propto \mu^{-\alpha+3/2}$ for large values of $\mu$, with $\al\in[2,3]$.  The probability,  $\mathcal{P}_{\text{obs}}(>\mu)$ is presented in Fig.\,\ref{PmuInt} for different values of $D_{\text{obs}}$.  From this scaling, it follows that the mean of the magnification at small observed distances diverges at large $\mu$. This  is due to the presence of rare but extremely highly magnified sources, i.e. sources aligned along the lens-observer direction for which the magnification $\mu\rightarrow\infty$, see eq.\,(\ref{e:muy}).\footnote{It also follows that, for any fixed value of $D_{\text{obs}}$ the variance of the distribution diverges as $\mu^{-\alpha+7/2}$. We stress that this is a general property of the probability distribution of magnification (for different observed redshift), independent on the details of the distribution of sources. Moreover, it applies also to the case of strongly lensed sources of electromagnetic radiation. Also in that case, the variance of the distribution of magnification (for different observed redshift) is divergent. }  Physically, the magnification tends to a finite value and, as mentioned above, the presence of the divergence is due to the geometric optics approximation, which breaks down in the vicinity of caustics, where wave effects need to be taken into account. Nevertheless,  this indicates that the distribution has a significant tail with large magnification. That the mean of the magnification diverges only for small observed distances is due to the fact that highly magnified sources appear to have small distances, $D_{\text{obs}}=D_s/\sqrt{\mu}$. In Fig.~\ref{mu2} and in the left panel of Fig.\,\ref{PmuInt} we present the results for the average amplification as a function of cosmological and observed redshift of a source\footnote{A cut-off in $\mu$ has been introduced to compute the mean for sources at low distances $D_{\text{obs}}<1000$ Mpc. We checked that our cut-off choice is such that increasing its value does not affect significantly the result.}. The average magnification as a function of the observed redshift is very close to $1$ for low redshift and it starts growing from $z_{\text{obs}}=0.5$, reaching the maximum value of $40$ at a redshift $0.8$ where we stop seeing sources, i.e. where the horizon of the observatory (computed allowing for the presence of magnification) lies.

In Section \ref{s:ana} we have analyzed the dependence of the  magnification of detected sources on the details of the population of sources, i.e. on the distribution of masses of objects in binary systems, using a simple analytic model that captures the main features. This analysis, valid for low redshift sources, shows that the average magnification  depends (mildly) on the details of the source population, but that the qualitative picture of Fig.\,\ref{mu2} is expected to be valid independent of the details of the astrophysical model chosen. The results for the average magnification as a function of cosmological redshift that we find with the analytic model is in very good agreement with the result of the numerical analysis, as can be seen by comparing Fig.\,\ref{An1} with the left panel of Fig.\,\ref{mu2}.

Let us finally derive some consequences for LIGO-Virgo O1 and O2 observations. The 10 observed BBH events have distances $D_{\text{obs}}\in [320,2750]$ Mpc. From the right panel of  Fig.\,\ref{PmuInt} it follows that (for our modeling) the probability that one of them has been magnified with magnification of 5 or bigger is $\mathcal{P}_{\text{obs}}(>5)\sim 0.01$. However, the probability that the magnification of one of them is bigger than 50 is not much smaller, namely $\mathcal{P}_{\text{obs}}(>50)\sim 0.005$.\\

Summarizing, in this article  we have presented a general framework to study lensing of GW by galaxies with a statistical approach. The framework is flexible and can be adapted to study strong and weak lensing of sources in any frequency band, and it can be applied to any detector network. We have presented quantities such as the probability of magnification as a function of cosmological or observed redshift, which do not depend on instrumental details, but only on the lens model. We have then presented results for the probability that a given observed event has been magnified by a given amount. This quantity depends on the details of the observatory considered (on its sensitivity curve).  Using an analytical model we have tested that the qualitative result for the distribution of magnification as a function of source redshift is solid under variations of the redshift and mass distributions of the source population.  In this work we have applied our formalism to the case of the LIGO-Virgo detectors. Future work will be dedicated to a detailed study of the ET and LISA cases, using the framework presented here.

\section*{Acknowledgements}
It is a pleasure to thank Michele Maggiore for helpful discussions. We also thank Paul Torrey for providing us with the analytic fitting formula of \cite{2015MNRAS.454.2770T}.  We are very grateful also to XiKai Shan who pointed out a relevant error in the first version of this paper.
This project has received funding from the European Research Council (ERC) under the European Union's Horizon 2020 research and innovation program (grant agreement No 693024) and from the Swiss National Science Foundation.

\FloatBarrier
\newpage

\appendix

\section{Strong lensing: notation and basic quantities}\label{AppA}

The lens equation is
\be
\bbt=\bth-\bal(\bth)\,,
\ee
where $\bal$ is the deflection angle.
We consider an  axi-symmetric lens with dimensionless surface density 
\be
\ka(\th) =\frac{\Si(|\bth|)}{\Si_{c}} \,, \qquad \th =|\bth| \,,
\ee
where we will define the critical surface density, $\Si_{c}$, conveniently such that $\ka$ becomes the convergence. Here $\bth$ is a 2-dim angle or a dimensionless variable on the surface normal to the line of sight, and the surface density is given in terms of the 
mass density $\rho$ as
\be
\Si(\th) = \int_{-\infty}^\infty d\ell \rho(\ell, d_{\ell}\th) \,.
\ee
Axi-symmetry implies that $\Si$ depends only on 
$\th= |\bth|$ and $d_{\ell}$ is the angular diameter distance to the center of the lens. The 2-dimensional deflection angle is then given by
\be
\bal(\bth) = \frac{1}{\pi}\int d^2\th' \frac{\bth-\bth'}{|\bth-\bth'|^2}\ka(\th')\,.
\ee
Integration yields 
\be\label{e:al1}
\bal(\bth) =  \frac{\bth}{\th}\al(\th) \,, \qquad \al(\th)= \frac{2}{\th}\int_0^\th d\th'\th'\ka(\th') \,.
\ee 
The time delay between two images  seen at positions $\bth^{(1)}$ and $\bth^{(2)}$ is given by 
\be
\Delta t=  t_1-t_2=(1+z_{\ell})\frac{d_s d_{\ell}}{d_{\ell s}}\left[\Phi(\bth^{(1)}\,, \bbt)-\Phi(\bth^{(2)}\,, \bbt)\right]\,,
\ee
where the Fermat potential $\Phi$ is
\be
\Phi(\bth,\bbt)=\frac{1}{2}(\bth-\bbt)^2-\psi(\bth)\,,
\ee
and $\psi(\bth)$ is the (dimensionless) lensing potential given by
\be
\psi(\bth) =\int d^2\th'\ka(\th')\log\left(\frac{|\bth-\bth'|}{\th_0}\right) + \mbox{const.} ~\,,
\ee 
defined such that $\bal=\bnabla\psi$.

In the axi-symmetric case, the lens equation reduces to the scalar equation 
\be
\beta=\theta-\alpha(\theta)\,.
\ee
In the limit $\beta\rightarrow 0$ the lensed
image forms a ring with radius $\theta_{\text{E}}$, solution of
\be
\theta_{\text{E}}-\alpha(\theta_{\text{E}})=0\,.
\ee

\section{Fit to the Illustris simulation}\label{app:Illustris}

In the following table we list the coefficients of the fitting formula used in Sec.\,\ref{catalogues} to describe the evolution of galaxy as a function of redshift and velocity dispersion \cite{Paul}. 

\begin{table}[ht!]
\centering
\begin{tabular}{c|ccc}
$i$& $i=0$ &  $i=1$&  $i=2$  \\
 \hline
 $A_i$& 7.391498 &5.729400  &  -1.120552  \\
 $\alpha_i$ &-6.863393  &  -5.273271&  1.104114  \\
$\beta_i$ &2.852083  &1.255696  &  -0.286638  \\
 $\gamma_i$ &  0.067032& -0.048683 &  0.007648\\
 \hline
\end{tabular}
\end{table}

\FloatBarrier 

\section{Analytic derivation of the average magnification}\label{appendixMu}

In this appendix we present in detail the analytic derivation of the average magnification as a function of the cosmological redshift of a source, using the results for the source distribution presented in section \ref{an:sources}. 

\subsection{Average magnification as a function of the limiting signal to noise}

We assume that the distribution of sources per unit of signal to noise and redshift s known. Our starting point is the following expression for the number density of sources as a function of signal to noise and redshift 
\be\label{NI2}
N(\rho,z_s)= \left\{\begin{array}{ll} N_{\rho}(z_s)\rho^{-\gamma}  & \mbox{for } \rho_1(z_s)<\rho<\rho_2(z_s)\,,\\
 0  & \mbox{otherwise,}\end{array} \right.
\ee
where $\gamma$ depends on the mass distribution for the two objects in the binary system as explained in the main text and $\rho_{1}\equiv \rho_{\text{min}}$, $\rho_2\equiv\rho_{\text{max}}$ correspond to the minimum and maximum value of the SNR in a given redshift bin and are defined in eqs.\,(\ref{min}) and (\ref{max}). 
The number of sources which we see from redshift $z_s$ with our limiting SNR $\rho_{\lim}$ if magnified by $\mu$ is
\begin{align}\label{e:Nlim}
N(\rho_{\lim},z_s,\mu)&= N_{\rho}(z_s)\int_{\alpha_{\lim}}^{\rho_2(z_s)} d\rho\,\rho^{-\gamma}\nn\\
&=\frac{N_{\rho}(z_s)}{\gamma-1}\left[\left(\frac{1}{\alpha_{\lim}}\right)^{\gamma-1}-\left(\frac{1}{\rho_2(z_s)}\right)^{\gamma-1}\right] \,,
\end{align}
where $\alpha_{\lim}=\text{max}\left(\rho_{\lim}/\sqrt{\mu}\,, \rho_1(z_s)\right)$. We assume $\rho_{\lim}/\sqrt{\mu}<\rho_2(z_s)$. If this is not the case, $N(\rho_{\lim},z_s,\mu)=0$ . We  assume here that $\gamma \neq 1$ but the case $\gamma=1$ can be easily worked out following steps similar to the ones performed below. The result without magnification is simply $N(\rho_{\lim},z_s,1)$. The total number of sources which we see is therefore
\be
N(\rho_{\lim},z_s)=\int_1^\infty d\mu p(\mu,z_s)N(\rho_{\lim},z_s,\mu)\,.
\ee
With a bit of algebra this can be rewritten as
\bea
N(\rho_{\lim},z_s) &=&\frac{N_{\rho}}{\gamma-1}(\rho_1^{-\gamma+1}-\rho_2^{-\gamma+1}) +\frac{N_{\rho}}{\gamma-1}\Bigg[\rho_2^{-\gamma+1}\int_1^{(\rho_{\text{lim}}/\rho_2)^2}d\mu\,p(\mu\,, z_s) \hspace*{2.2cm}\nonumber\\  \label{e:Mlim}
&& \hspace*{0.3cm} -\rho_1^{-\gamma+1}\int_1^{(\rho_{\text{lim}}/\rho_1)^2}d\mu\,p(\mu\,, z_s)+\int_{(\rho_{\text{lim}}/\rho_2)^2}^{(\rho_{\text{lim}}/\rho_1)^2}d\mu\,p(\mu\,, z_s)\left(\frac{\sqrt{\mu}}{\rho_{\text{lim}}}\right)^{\gamma-1}\Bigg]\,,
\eea
 where the redshift dependence in $\rho_{1,2}$ and $N_{\rho}$ is understood. 
We eventually want to compute the average magnification. For this  we need to compute 
\be\label{e:Mlim}
M(\rho_{\lim},z_s)\equiv \int_1^\infty d\mu \,\mu\, p(\mu,z_s)N(\rho_{\lim},z_s,\mu)\,,
\ee
which can be done with similar steps as above yielding 
\bea
M(\rho_{\lim},z_s) &=&\frac{N_{\rho}}{\gamma-1}(\rho_1^{-\gamma+1}-\rho_2^{-\gamma+1})\bar{\mu} +\frac{N_{\rho}}{\gamma-1}\Bigg[\rho_2^{-\gamma+1}\int_1^{(\rho_{\text{lim}}/\rho_2)^2}d\mu\,\mu\,p(\mu\,, z_s) \hspace*{2.2cm}\label{e:Mlim}\nn\\
&& \hspace*{-0.3cm} -\rho_1^{-\gamma+1}\int_1^{(\rho_{\text{lim}}/\rho_1)^2}d\mu\,\mu\,p(\mu\,, z_s)+\int_{(\rho_{\text{lim}}/\rho_2)^2}^{(\rho_{\text{lim}}/\rho_1)^2}d\mu\,\mu\,p(\mu\,, z_s)\left(\frac{\sqrt{\mu}}{\rho_{\text{lim}}}\right)^{\gamma-1}\Bigg]\,,
\eea
where $\bar\mu$ depends on the source redshift and is defined as
\be
\bar{\mu}(z_s)=\int_1^{\infty} d\mu\,\mu p(\mu,z_s)\,.
\ee
The average magnification is now given by  the ratio of (\ref{e:Mlim}) over (\ref{e:Nlim}) 
\be
\langle \mu\rangle(z_s)=M(\rho_{\lim},z_s)/N(\rho_{\lim},z_s)\,.
\ee
Completely analogous steps can be performed working with the number density as a function of intrinsic luminosity, as we show in the following. 

\subsection{Average magnification as a function of the  limiting luminosity}

Here we derive  the analytic expression of the amplification as a function of redshift assuming that we know the distribution of sources per unit of intrinsic luminosity and redshift. Our starting point is the following expression for the number density of sources as a function of luminosity and redshift 
\be\label{NI}
N(L,z_s)= \left\{\begin{array}{ll} N_L(z_s)L^{-\beta}  & \mbox{for } L_1(z_s)<L<L_2(z_s)\,,\\
 0  & \mbox{otherwise,}\end{array} \right.
\ee
where $\beta$ depends on the mass distribution for the two objects in the binary system as explained in the main text. In eq.\,(\ref{NI}), $L_{1}$ and $L_2$ correspond to the minimum and maximum value of the luminosity in a given bin, and can be easily related to the minimum and maximum signal to noise (considering a given detector) making use of eqs.\,(\ref{Finn0}) and (\ref{Finn}). 
The number of sources which we see from redshift $z_s$ with our limiting flux $F_{\lim}\propto \rho_{\lim}^2$ when magnified by $\mu$ is
\begin{align}\label{num}
N(F_{\lim},z_s,\mu)&= \int_{\alpha_{\lim}}^\infty dL\,N_L(z_s)L^{-\beta}\nn\\
&=\frac{1}{\beta-1}N_L(z_s)\left[\left(\frac{1}{\alpha_{\lim}}\right)^{\beta-1}-\left(\frac{1}{L_2(z_s)}\right)^{\beta-1}\right] \,,
\end{align}
where $\alpha_{\lim}=\text{max}\left(L_{\lim}/\mu\,, L_1(z_s)\right)$. We are here assuming that $\beta \neq 1$ but the case $\beta=1$ can be easily worked out following steps similar to the ones proposed below. Without magnification this is simply $N(F_{\lim},z_s,1)$. The total number of sources which we see is therefore
\be
N(F_{\lim},z_s)=\int_1^\infty d\mu p(\mu,z_s)N(F_{\lim},z_s,\mu)\,.
\ee
With a bit of algebra this can be rewritten as
\begin{align}
N(F_{\lim},z_s)=&\frac{N_L}{\beta-1}(L_1^{-\beta+1}-L_2^{-\beta+1})+\frac{N_L}{\beta-1}\left[L_2^{-\beta+1}\int_1^{L_{\text{lim}}/L_2}d\mu\,p(\mu\,, z_s)\right. \nn\\
&\left.-L_1^{-\beta+1}\int_1^{L_{\text{lim}}/L_1}d\mu\,p(\mu\,, z_s)+\int_{L_{\text{lim}}/L_2}^{L_{\text{lim}}/L_1}d\mu\,p(\mu\,, z_s)\left(\frac{\mu}{L_{\text{lim}}}\right)^{\beta-1}\right]\,, \label{e:Nlim2}
\end{align}
 where to make the notation compact the redshift dependence in $L_{1,2}$ and $N_L$ is understood. 
We eventually want to compute the average magnification. For this  we need to compute 
\be\label{den}
M(F_{\lim},z_s)\equiv \int_1^\infty d\mu \,\mu\, p(\mu,z_s)N(F_{\lim},z_s,\mu)\,,
\ee
which can be done with similar steps and gives 
\bea
M(F_{\lim},z_s) &=&\frac{N_L}{\beta-1}(L_1^{-\beta+1}-L_2^{-\beta+1})\bar{\mu} +\frac{N_L}{\beta-1}\Bigg[L_2^{-\beta+1}\int_1^{L_{\text{lim}}/L_2}d\mu\,\mu\,p(\mu\,, z_s) \hspace*{2.2cm}\nonumber\\  \label{e:Mlim2}
&&  -L_1^{-\beta+1}\int_1^{L_{\text{lim}}/L_1}d\mu\,\mu\,p(\mu\,, z_s)+\int_{L_{\text{lim}}/L_2}^{L_{\text{lim}}/L_1}d\mu\,\mu\,p(\mu\,, z_s)\left(\frac{\mu}{L_{\text{lim}}}\right)^{\beta-1}\Bigg]\,,
\eea
where 
\be
\bar{\mu}=2p_1\int_1^{\infty} \frac{d\mu\,\mu}{(\mu-1)^3}e^{-\frac{p_1}{(\mu-1)^2}}\,,
\ee
and the average magnification is just given by  (\ref{e:Mlim2}) over (\ref{e:Nlim2}) 
\be
\langle \mu\rangle(z_s)=M(F_{\lim},z_s)/N(F_{\lim},z_s)\,.
\ee

 \bibliographystyle{JHEP}
\bibliography{refs}

\end{document}